\documentclass[12pt,a4paper,twoside,dvips]{article}
\usepackage{a4p}
\usepackage{cite,mcite}
\usepackage{physics}
\usepackage{l3_title}
\usepackage{ifthen}
\usepackage{Lep}
\usepackage{graphicx}
\usepackage{amssymb}
\usepackage{pennames}
\journalname{Phys. Lett. B}
\date{July 16, 2003}
\preprint{2003-047}
\graphicspath{{.}}
\Lep{1}
\newlength{\capindent}
\newlength{\capwidth}
\newlength{\figwidth}
\newcommand{\icaption}[2][!*!,!]{\hspace*{\capindent}%
  \begin{minipage}{\capwidth}
    \ifthenelse{\equal{#1}{!*!,!}}%
      {\caption{#2}}%
      {\caption[#1]{#2}}
  \end{minipage}}
\def\abqq{A^{B}_{12}}

\def\asqq{A^{S}_{12}}

\def\B{B}

\def\epem{\mathrm{e^{+}e^{-}}}

\def\etal{\it et~al.}

\def\pp{\mathrm{p\bar{p}}}
\def\pt{\mathrm{p_{T}}}
\def\qg{\mathrm{qg}}

\def\qbar{\mathrm{\bar{q}}}
\def\qqbar{\mathrm{q\bar{q}}}
\def\rs{\mathrm{\sqrt{s}}}
\def\S{S}
\def\sQ{\mathrm{\sigma_{Q}}}

\def\Pnj{\prod_{j=1}^n P_j}

\def\sst{\scriptscriptstyle}

\def\ra{\rightarrow}

\def\be{\begin{equation}}
\def\ee{\end{equation}}
\def\bea{\begin{eqnarray}}
\def\eea{\end{eqnarray}}

\def\csp{\textsc{Csp}~}  
\def\cop{\textsc{Cop}~}

\begin{document}
\begin{titlepage}

\title{Search for \\ Colour Singlet and 
       Colour Reconnection Effects \\ in Hadronic Z Decays
       at LEP}

\author{The L3 Collaboration}

\begin{abstract}
\vskip 0.3cm
A search is performed in symmetric 3-jet hadronic Z-decay events  for evidence 
of colour singlet production or colour reconnection effects. Asymmetries in 
the angular separation of particles are found to be sensitive indicators of 
such effects. Upper limits on the level of colour singlet production or of
colour reconnection effects are established for a variety of models.
\end{abstract}

\submitted

\end{titlepage}

\normalsize

\section{Introduction}

\begin{figure}[htbp]
\begin{center}
\includegraphics[width=.45\textwidth]{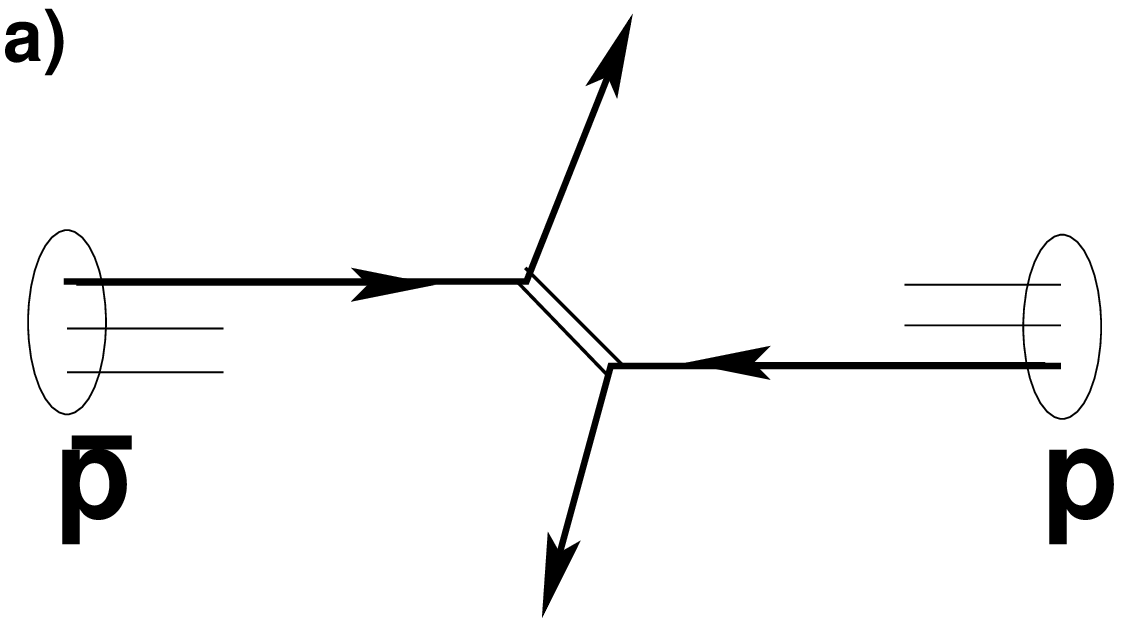} \hfill
\includegraphics[width=.49\textwidth]{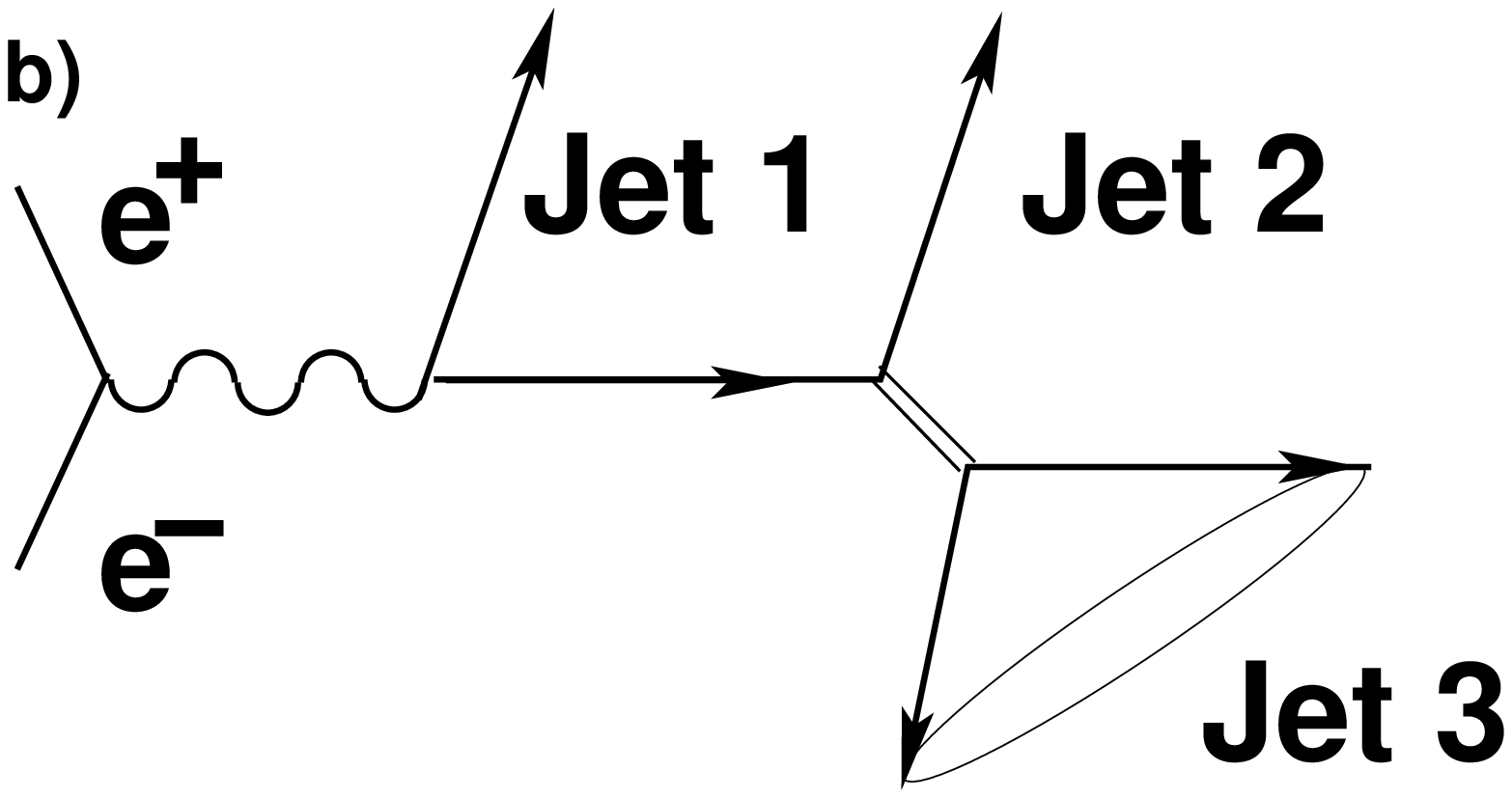}

\vspace*{5mm}
\includegraphics[width=.46\textwidth]{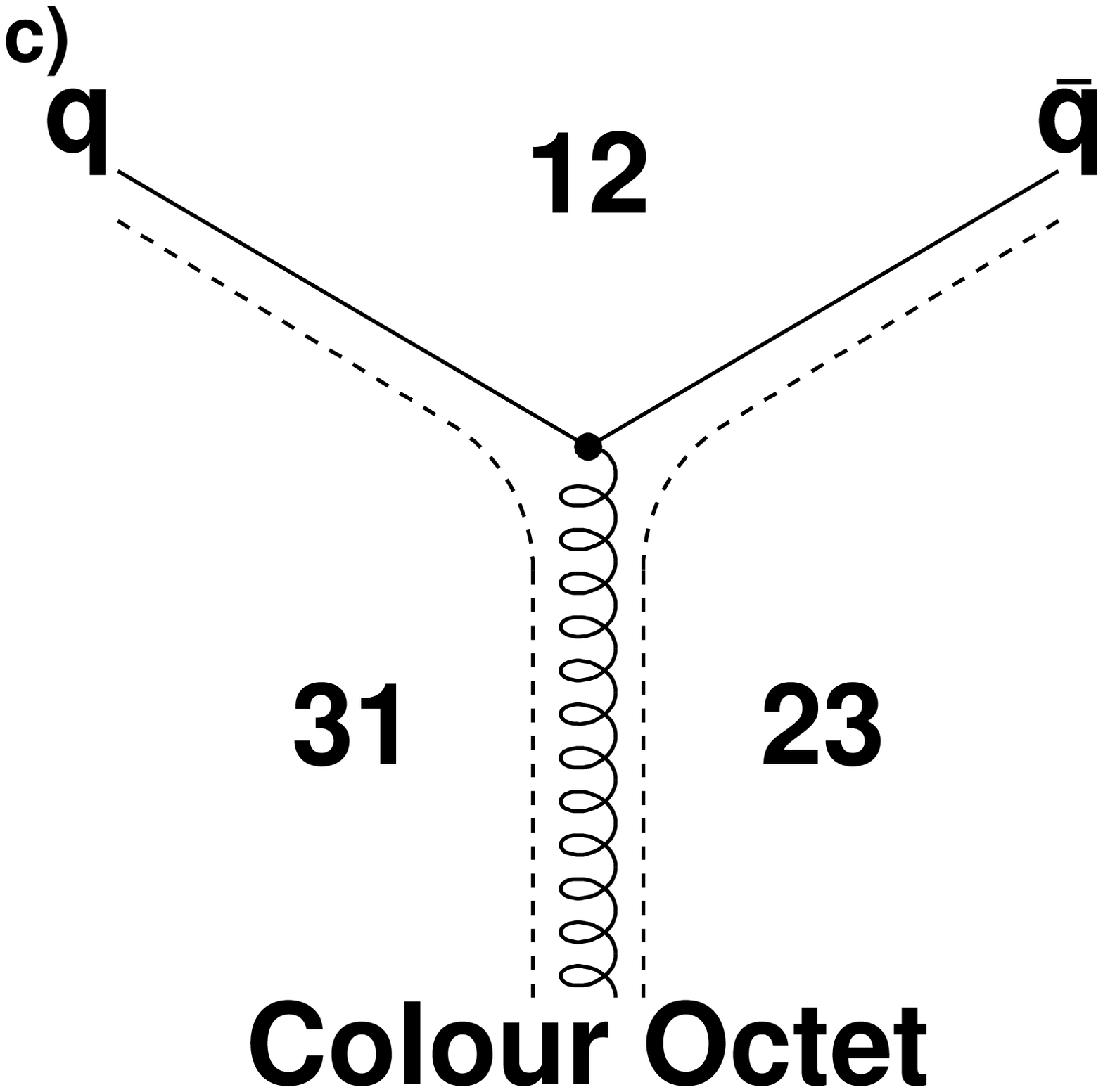} \hfill
\includegraphics[width=.46\textwidth]{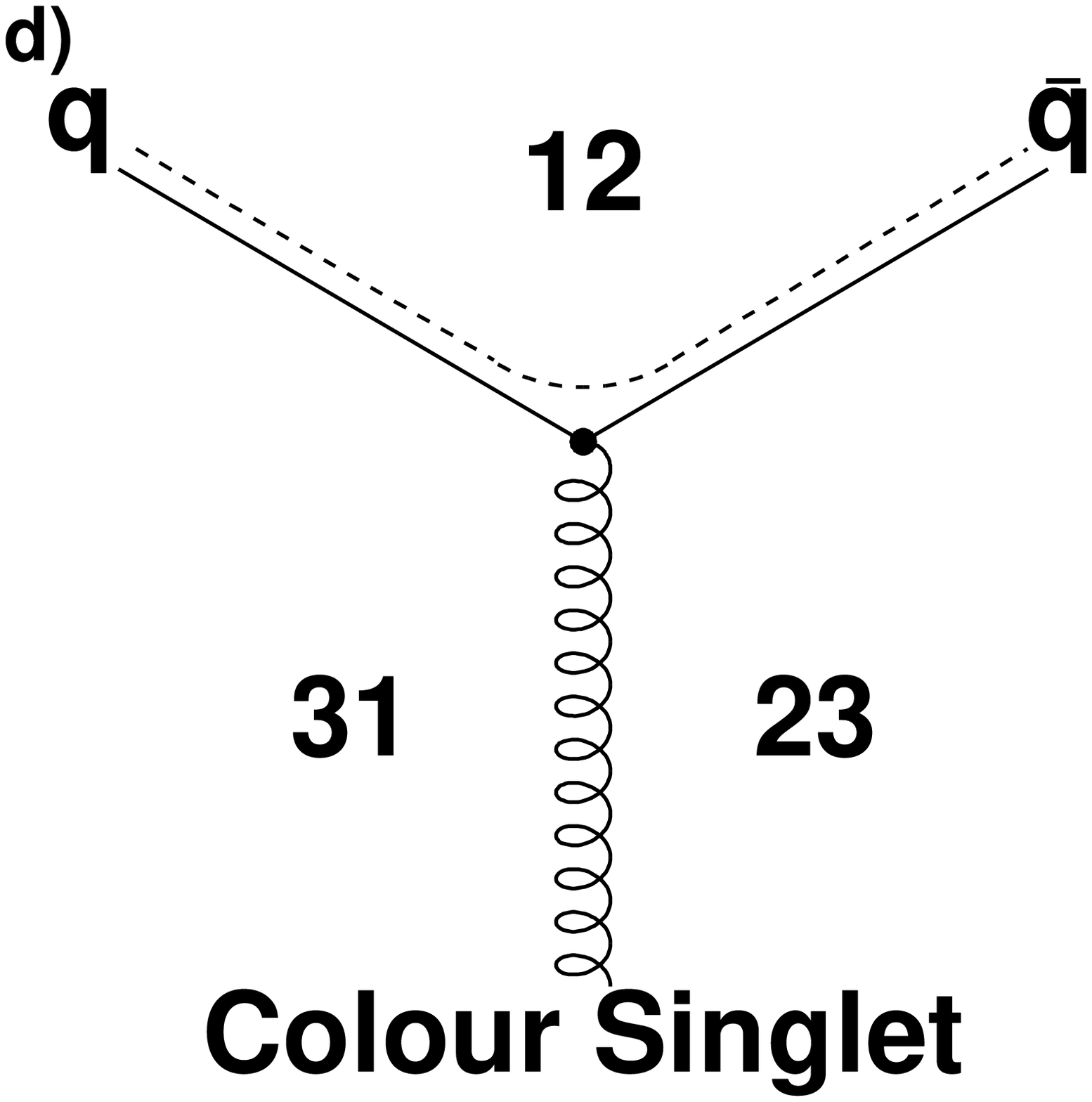}

\vspace*{5mm}
\includegraphics[width=.38\textwidth]{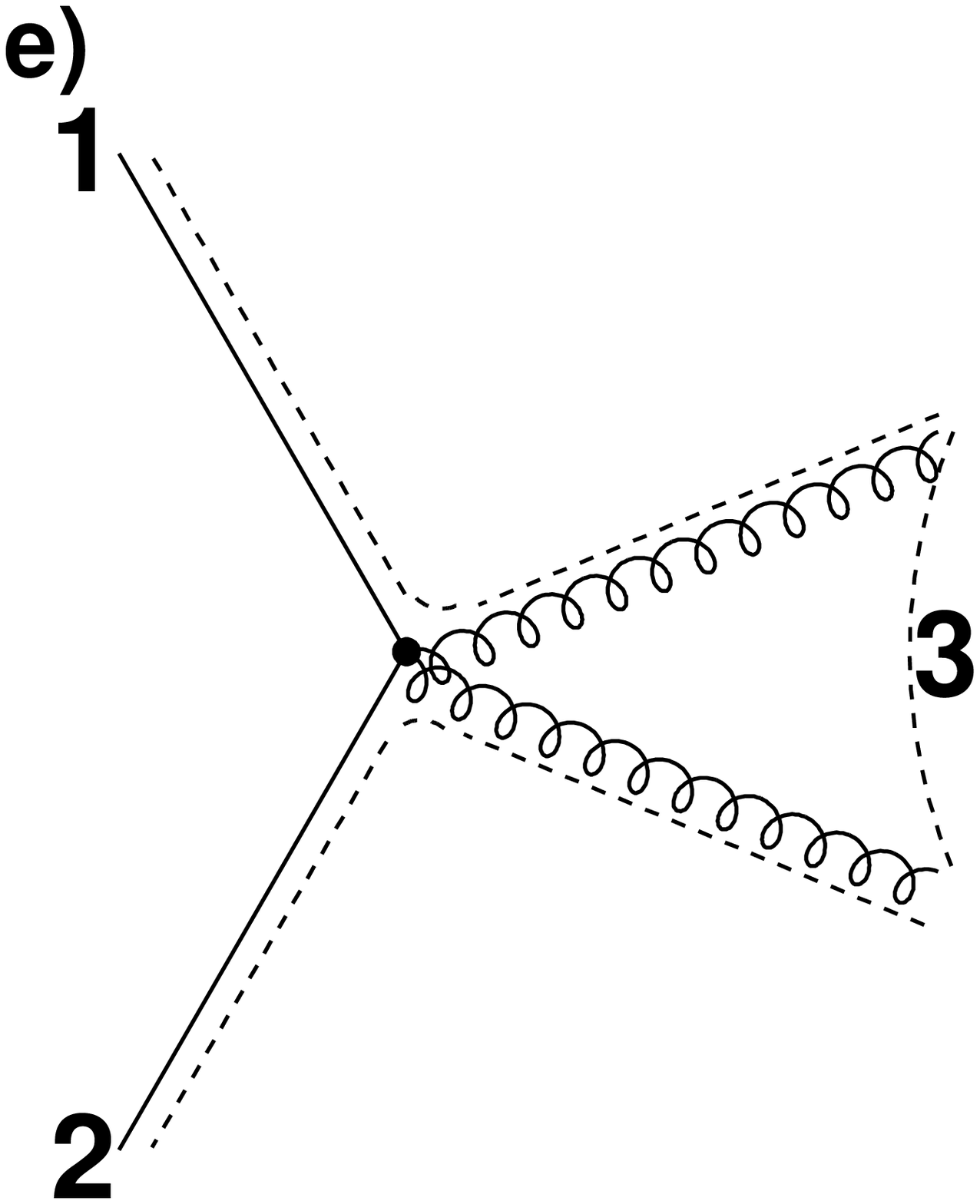} \hspace*{5mm}
\includegraphics[width=.38\textwidth]{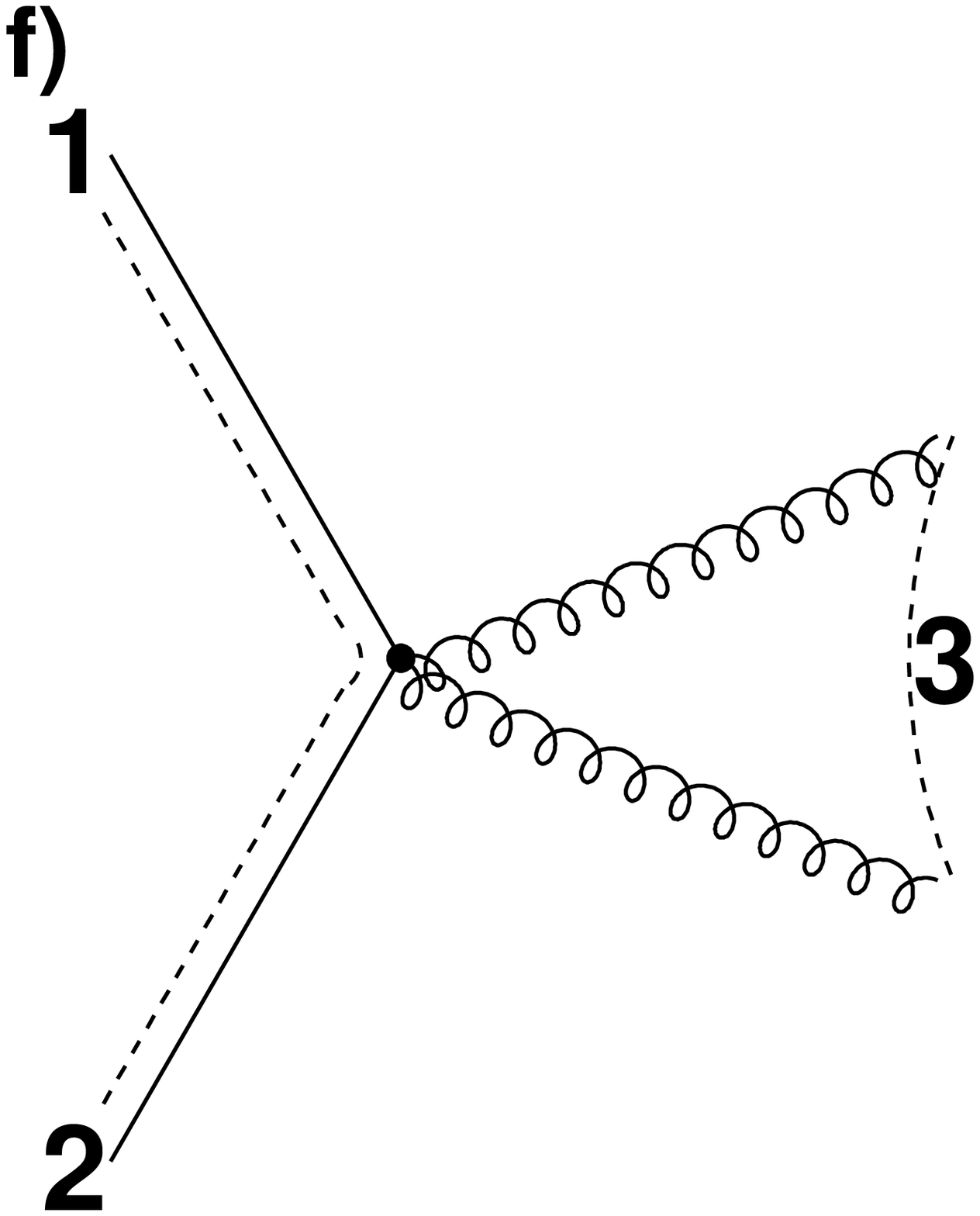}
 \caption[Schematic diagram of colour exchange. Colour singlet propagators
          are indicated by double lines in a) $\pp$ and in b) $\epem$ 
          reactions. The $\epem$ diagram is derived by crossing the incoming 
          quark line in the $\pp$ diagram. Colour flow is shown by dashed 
          lines for c) \cop and d) \csp in 3-jet events from $\epem$ 
          annihilation and also e) without and f) with colour reconnection.]
         {Schematic diagram of colour exchange. Colour singlet propagators
          are indicated by double lines in a) $\pp$ and in b) $\epem$ 
          reactions. The $\epem$ diagram is derived by crossing the incoming 
          quark line in the $\pp$ diagram. Colour flow is shown by dashed 
          lines for c) \cop and d) \csp in 3-jet events from $\epem$ 
          annihilation and also e) without and f) with colour reconnection.}
\label{fig:cross}
\end{center}
\end{figure}

The term `rapidity gaps' denotes regions of angular phase space devoid of 
particles. They are expected in low-$p_T$ diffractive processes, where separate
colour singlet hadronic systems are produced, well separated in phase space
and associated with either the target or projectile particles. Events 
containing large rapidity gaps, attributed to colour singlet exchange or 
colour reconnection effects, are also observed, in association with high-$p_T$ 
jet production, at {\textsc{Hera}}~\cite{hera} and at the 
{\textsc{Tevatron}}~\cite{tevatron}. As shown in Figures \ref{fig:cross}a and
\ref{fig:cross}b, by crossing symmetry, similar gaps may be expected in 
three-jet hadronic Z decays at LEP. The corresponding diagram has four final 
state partons, but, because of its generally low energy, the two jets 
associated with the colour singlet object are typically unresolved. 

Large rapidity gaps are observed in $\simeq$10$\%$ and 1--2$\%$ of events 
with two high-$p_T$ jets at {\textsc{Hera}} and at the {\textsc{Tevatron}},
respectively. In  electron-hadron or hadron-hadron  collisions particles
produced by the spectator partons of the underlying event frequently  
destroy large rapidity gaps associated with the primary hard scattering 
process. The associated `gap survival probability' is estimated~\cite{survprob}
to be about 20$\%$ at the {\textsc{Tevatron}}. An advantage of the Z decay 
study is the absence of this suppression factor, as there is no underlying 
event. A disadvantage is that the maximum possible size of the angular gap is 
smaller for Z decays compared to ep or p${\rm \overline{p}}$ collisions.
Particle and energy flow in the inter-jet regions have
been studied~\cite{qqglue} using three-jet events in $\epem$ annihilations.
These studies revealed that the region between two-quark jets have lower
particle and energy flows relative to na\"{\i}ve expectation from
independent fragmentation models. This was also observed in studies
which compared three-jet events with two-jet events having a hard photon in the
final state~\cite{qqgamma}.

The analysis presented in this Letter is performed with hadronic Z-decay
events recorded by the L3 detector~\cite{L3} using 75.14 pb$^{-1}$ of data 
from the 1994--1995 Z-pole data taking periods. In the new method presented 
here, a search is performed for gaps in angular phase space in symmetric 
3-jet hadronic Z-decay events. The method exploits the different particle 
flows between quark and antiquark jets and either the quark or antiquark jet 
and the gluon jet. After removing particles near to the jet cores, angular 
gaps between particles in the inter-jet regions are analysed, and various 
asymmetry variables are formed, as detailed below.

Studies of rapidity gaps in hadronic Z decays using as variable the
pseudo-rapidity of particles relative to the thrust axis were
previously reported~\cite{SLD}. A recent study~\cite{OPALrg} used the
axes of tagged gluon jets. The analysis presented in this Letter,
based on global event variables, is complementary to this study in the
sense that the jet cores are excluded from the analysis so as to
minimise fragmentation effects whereas previous analyses rather used
rapidity gaps as a tool to investigate the details of gluon
fragmentation~\cite{OPALrg}.

The present analysis extends the notion of comparing particle and
energy flow in the region opposite to the quark jets as well as in the
region between the two-quark jets by introducing new asymmetry
variables that are sensitive to the relative difference in colour flow
between all the inter-jet regions. 

A first application of these asymmetries is to exploit differences in
colour flow between events where colour singlet systems are produced
(\textsc{Csp}) and conventional gluon colour octet production
(\textsc{Cop}). As shown in Figure \ref{fig:cross}c, in \textsc{Cop},
colour flow is present between the qg and $\qbar$g gaps and is
inhibited by destructive interference in the $\qqbar$ gap. As shown in
Figure \ref{fig:cross}d, the colour string in the \textsc{Csp} is
drawn between the quark and the anti-quark so that an appreciable
colour flow occurs also in the $\qqbar$ gap.
 
A second application is to investigate colour reconnection (CR) effects.
Partons originating from a hard scattering process are eventually transformed
into hadrons and this hadronisation process requires specification of the 
colour flow pattern among the partons. In the simplest models, the colour flow 
associated with the final state partons is fixed during the hard scattering 
process. However, there may be subsequent rearrangement of the colour flow. 
At the perturbative level this requires the exchange of at least two gluons 
between the partons. Coloured strings, normally stretched between a quark and 
a gluon as shown in Figure \ref{fig:cross}e, can be rearranged in the
colour reconnection picture so as to create colour singlet
quark pair in association with a colour singlet gluon 
pair, whose colour strings then hadronise independently, as shown in Figure
\ref{fig:cross}f. To study CR effects, the \textsc{Gal}~\cite{rath} model 
as well as CR as implemented in \textsc{Ariadne} \cite{ar2} and \textsc{Herwig}
\cite{herwigcr} are considered.

Studies of the determination of the W boson mass using fully hadronic W-pair 
decays, indicate CR effects as the dominant source of theoretical systematic 
uncertainty. If the same CR algorithm is valid for both Z and W-pair decays, 
limits on the level of CR effects, established experimentally at the Z-pole,
can be used to constrain the systematic uncertainty on the W-mass 
determination. The present analysis is thus complementary to the direct 
measurement of CR effects in hadronic decays of W-pairs~\cite{DDWCR}.

\section{Event and Particle Selection}
 
Well balanced hadronic Z-decay events are selected by cuts on the number of 
calorimetric clusters with energy greater than 100 MeV, $N_{cluster}$, on  
the total energy observed in the calorimeters, $E_{vis}$, the energy imbalance 
along the beam direction, $E_{\parallel}$, and the energy imbalance in the 
plane perpendicular to the beam direction, $E_{\perp}$. The cut on the number 
of calorimetric clusters rejects low multiplicity events such as 
$\tau^-\tau^+$ final states. About two million hadronic Z-decay candidates are 
selected.

\begin{figure}[htbp]
  \begin{center}
    \includegraphics[width=.48\textwidth]{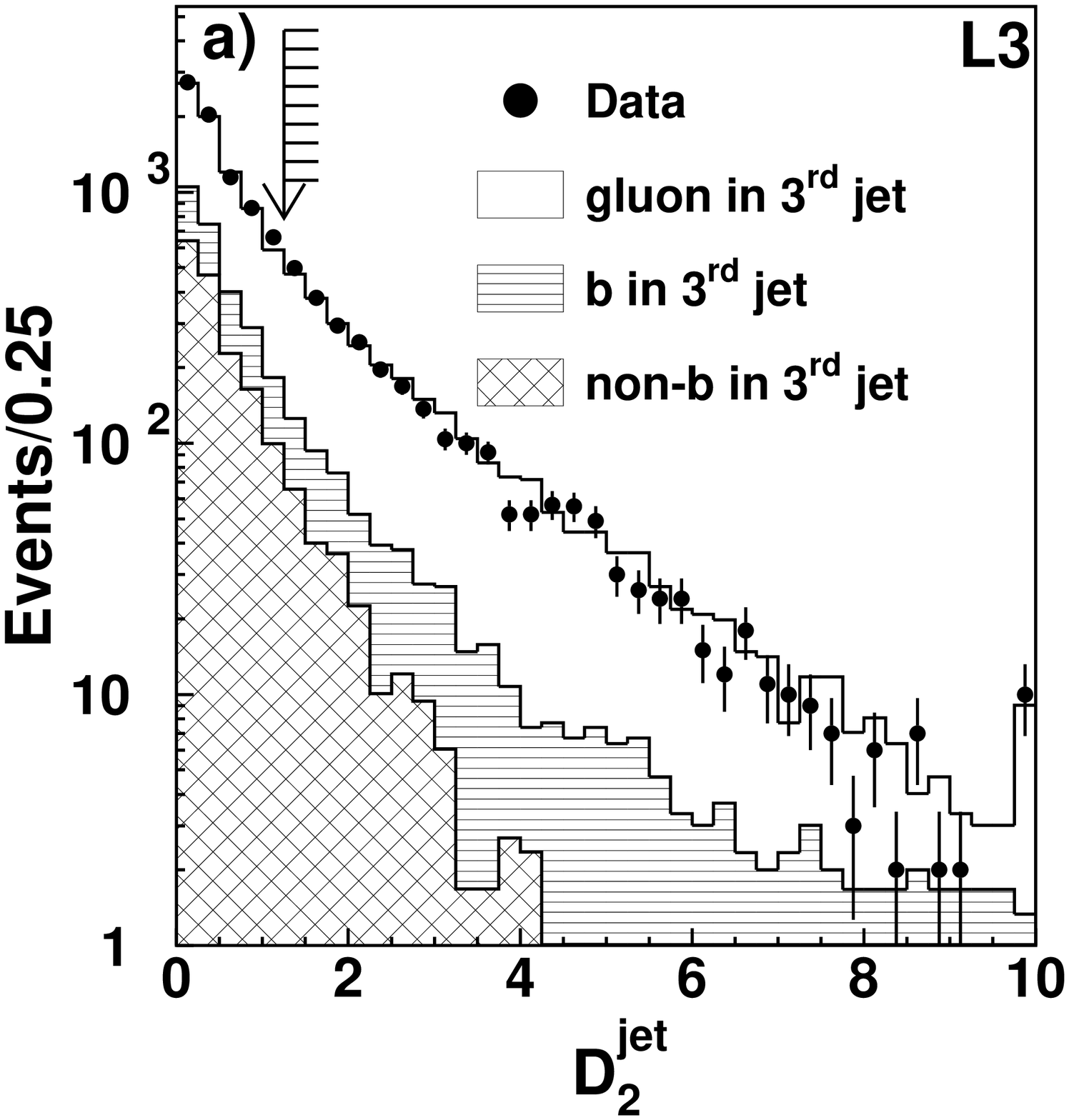}
    \includegraphics[width=.48\textwidth]{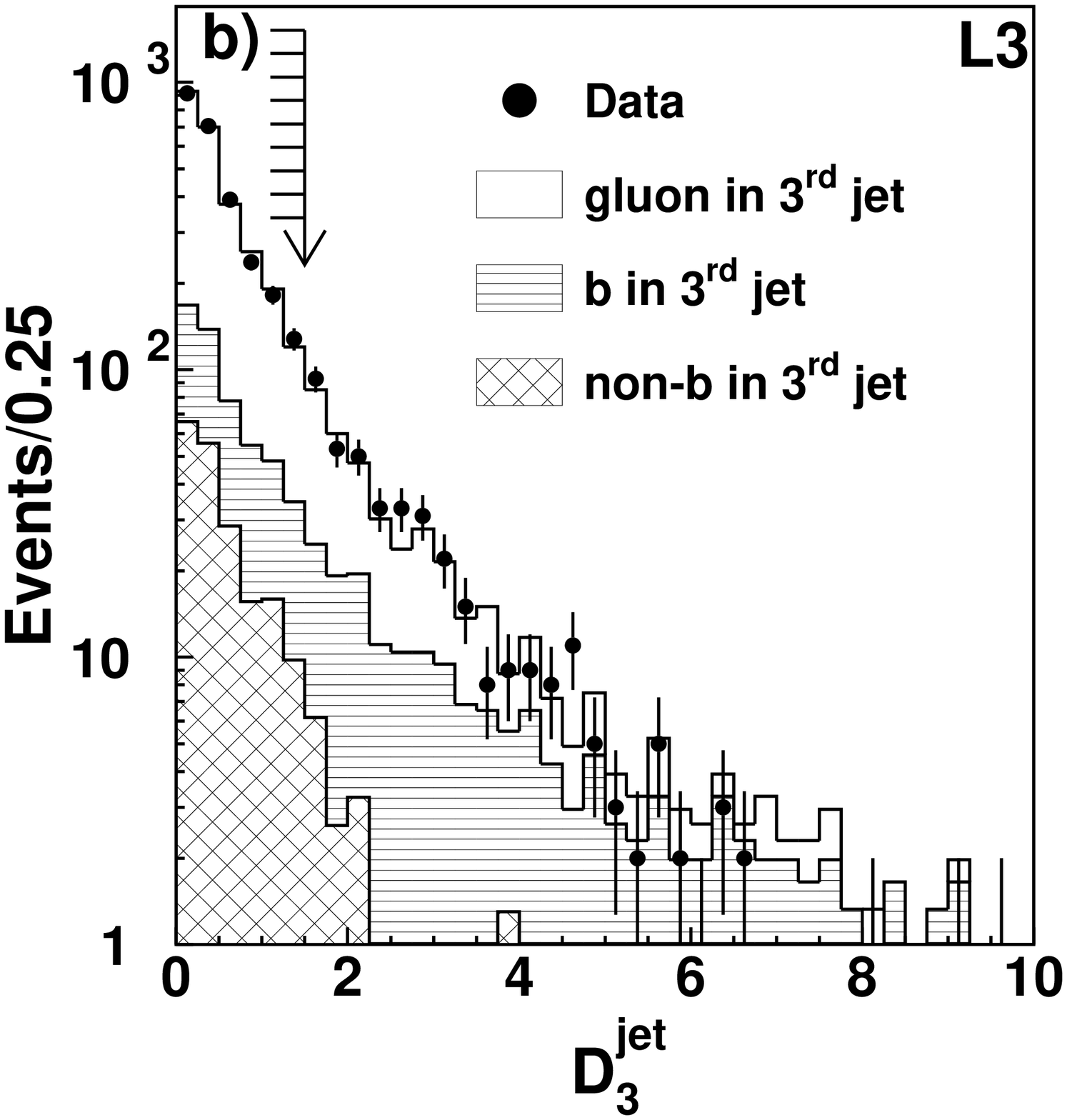}
  \caption[b-tag discriminant plots for energy-ordered jets: a) 
          for jet 2 and b) for jet 3. Vertical arrows represent the cuts.]
          {b-tag discriminant plots for energy-ordered jets: a) 
          for jet 2 and b) for jet 3. Vertical arrows represent the cuts.}
  \label{fig:btagj}
  \end{center}
\end{figure}

Symmetric three-jet events with a jet-jet angular separation of about 
$120^{\circ}$ are then selected using the \textsc{Jade} algorithm~\cite{jade}, 
with the jet resolution parameter set to 0.05. The angles between jets $i$ and 
$j$, $\phi_{ij}$, are required to be within $\pm 30^{\circ}$ of the symmetric 
topology. Using the selection criteria:
\[ N_{cluster} > 12,~~0.6 < E_{vis}/\sqrt{s} < 1.4,~~E_\parallel / E_{vis} 
< 0.40, \]
\[ E_\perp / E_{vis} < 0.40,~~~\phi_{12},\phi_{23}, \phi_{31} ~~ \in ~~ 
[90^{\circ}, 150^{\circ}], \]
where $\sqrt{s}$ is the centre-of-mass energy, about 70000 three-jet events 
are obtained.  In order to distinguish quark jets from gluon or colour singlet 
jets, the energy-ordered quark jets are tagged by cuts on the b-tag
discriminant\footnote{The jet b-tag discriminant of jet $i$, containing $n$ 
tracks, is defined as: $D^{jet}_i = -\log_{10}P$ where $ P  = P_i^n
\sum_{j=0}^{n-1}(-\ln P_i^n)^j/j!$ and $P_i^n~=~\Pnj$. Here, $P_j$ is the 
probability that the $j$th track in the jet originates at the primary vertex.}:
\[ D^{jet}_1 > 1.25,~~ D^{jet}_2 > 1.25~~ D^{jet}_3 < 1.5\] 
As shown in Figure \ref{fig:btagj}, these cuts strongly enhance the gluon
fraction in jet 3. This selection tags 2668 events with a gluon purity of 78\%.

To study the particle flow, calorimetric clusters are selected which satisfy
at least one of the following criteria:
 \begin{itemize}
  \item energy greater than 100 MeV in the electromagnetic calorimeter (ECAL) 
        and at least 900 MeV in the hadron calorimeter (HCAL); 
  \item energy greater than 100 MeV in the ECAL with a minimum of 2 crystals hit;
  \item energy greater than 1800 MeV  in the HCAL alone.
 \end{itemize}
These cuts reject noisy clusters and take into account different thresholds in 
the calorimeters. The distributions of the cluster multiplicity  with these
selection criteria show good agreement between data and Monte Carlo, with residual differences below 2.5\%.

\section{ Monte Carlo Samples}

The \textsc{Jetset} Parton Shower (PS) Monte Carlo program~\cite{jetset} is 
used to model \textsc{Cop}. Two simple models are used to simulate the 
expected colour flow in \textsc{Csp}: events of type $\qqbar\gamma$ are 
generated with a photon effective mass as in the gluon jet mass distribution. 
In the first model, \textsc{Cs1}, the photon is replaced by a boosted di-quark 
jet. In the second model, \textsc{Cs2}, the photon is replaced by a  gluon 
fragmenting independently. The total particle multiplicity for both these 
models agree with \textsc{Jetset} within $\pm 1$ unit.

For CR studies, the \textsc{Gal} model, implemented in the \textsc{Pythia} 
Monte Carlo program \cite{pythia}, uses a default value of 0.1 for the colour 
recombination parameter, $R_0$. This value is obtained~\cite{rath} by fitting 
the model to H1 data on the diffractive proton structure function. For this 
study the fragmentation parameters of the model are tuned to Z-decay data\footnote{The 
QCD models are tuned using several global event shape
distributions  at $\rs$ $\approx$ $m_{\rm Z}$: the
minor on the narrow side~\cite{mns}, the jet resolution parameter 
for the  transition from 2- to 3-jet in the \textsc{Jade}~\cite{jade} algorithm, the 
fourth Fox-Wolfram moment~\cite{fwm} and the charged particle multiplicity. 
For models implemented in \textsc{Pythia}, 
the tuned parameters are the QCD cut-off parameter, $\Lambda$, and the string 
fragmentation parameters, $b$ and $\sQ$, affecting longitudinal and transverse 
components of the hadron momenta. In \textsc{Herwig}, the QCD cut-off parameter
and the parameters controlling hadronisation, \textsc{clmax} (maximum cluster 
mass) and \textsc{clpow} (the power of the mass in the expression for the 
cluster splitting criterion) are tuned.} 
for three different values of the colour recombination parameter: $R_0$ = 0.05,
0.1, 0.2.  The \textsc{Ariadne} and \textsc{Herwig} generators, with and 
without CR, are also tuned to Z-decay data to determine their basic 
fragmentation parameters. 
The colour reconnection probability in \textsc{Herwig} is set to
its default value of 1/9~\cite{herwigcr}.
Similarly, the parameters affecting colour reconnection in
\textsc{Ariadne} are kept at their default values~\cite{ar2},
\textsc{para(26)=9} and \textsc{para(28)=0}.
 
\section{Inter-Jet Gap Asymmetries}

After selection of three-jet events, the particle momenta are
projected onto the event plane defined by the two most energetic jets.
In order to minimise the bias from jet fragmentation, particles in a
cone of $15^\circ$ half-angle about the jet axis direction are
excluded from the analysis.  The angles of the remaining particles are
measured in this plane with respect to the most energetic jet. In
order to achieve uniformity in the event-to-event comparison, these
angles are rescaled so as to align jets at 0$^\circ$, 120$^\circ$ and
240$^\circ$. This is achieved by scaling the angle of a particle to
its nearest jet by the ratio between 120$^\circ$ and the opening angle
of the two jets between which the particle is located.
       
\begin{figure}[htbp]
\begin{center}
\includegraphics[width=.48\textwidth]{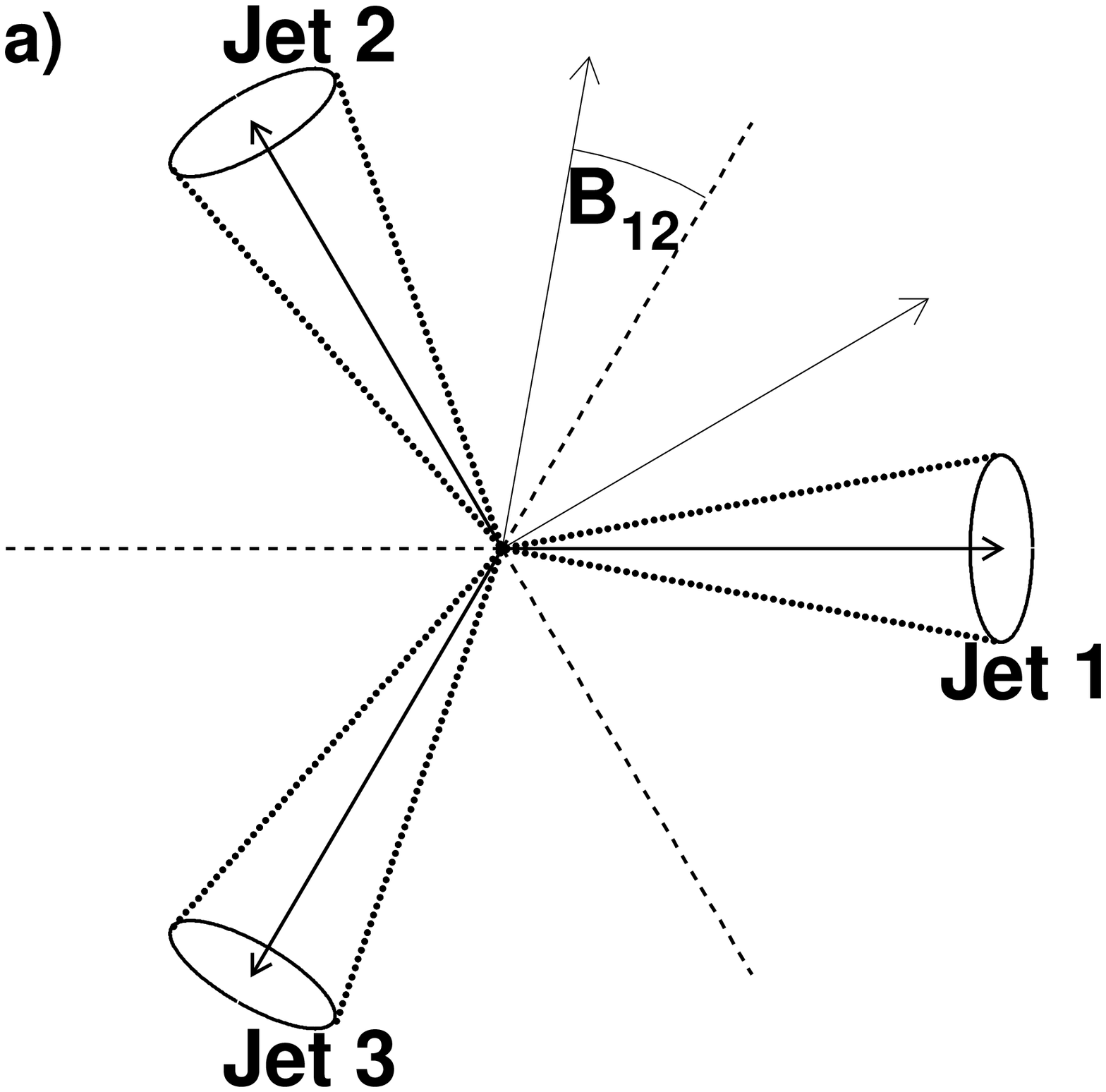} \hfill
\includegraphics[width=.48\textwidth]{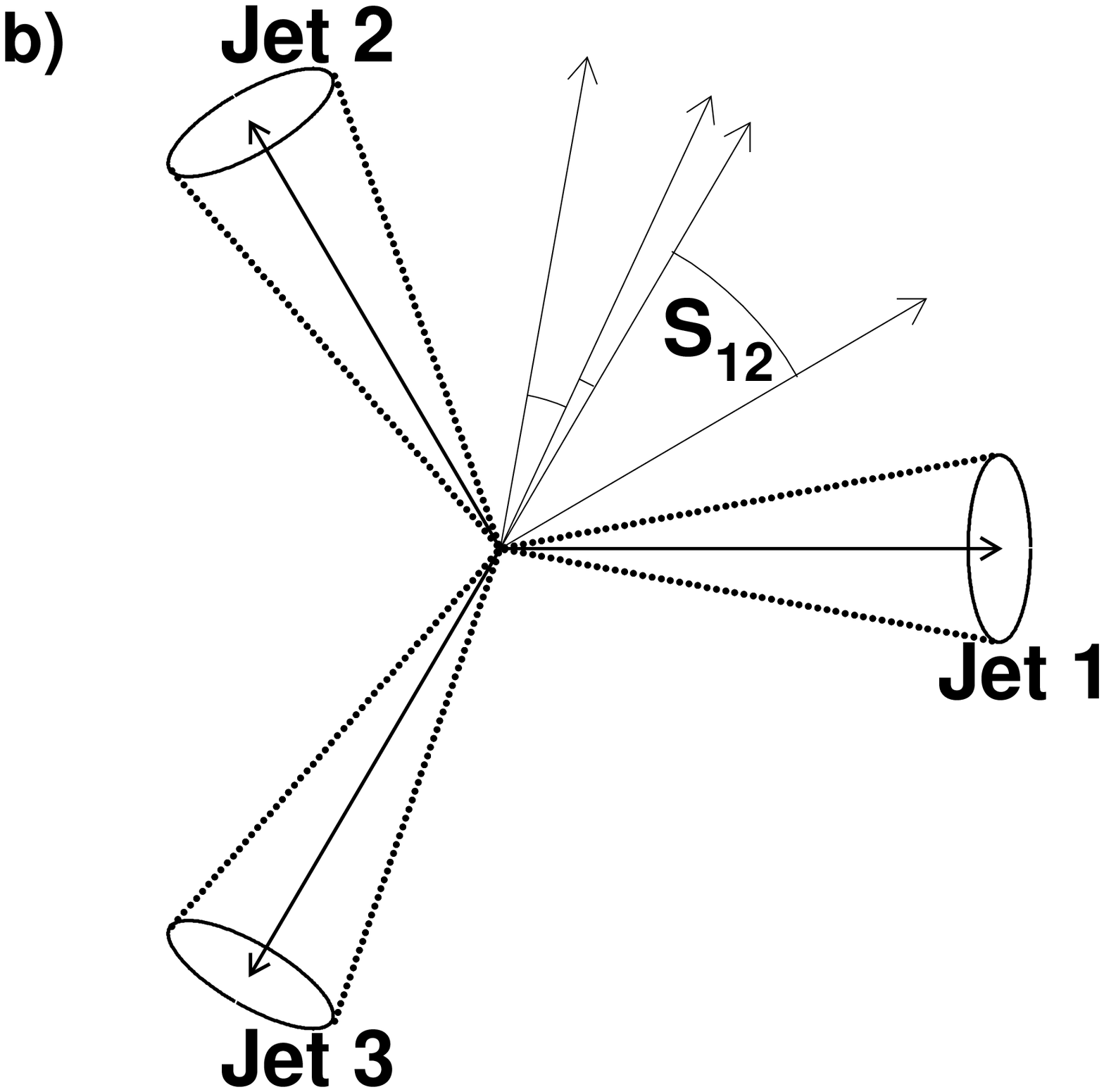}

\vspace*{10mm}
\includegraphics[width=.48\textwidth]{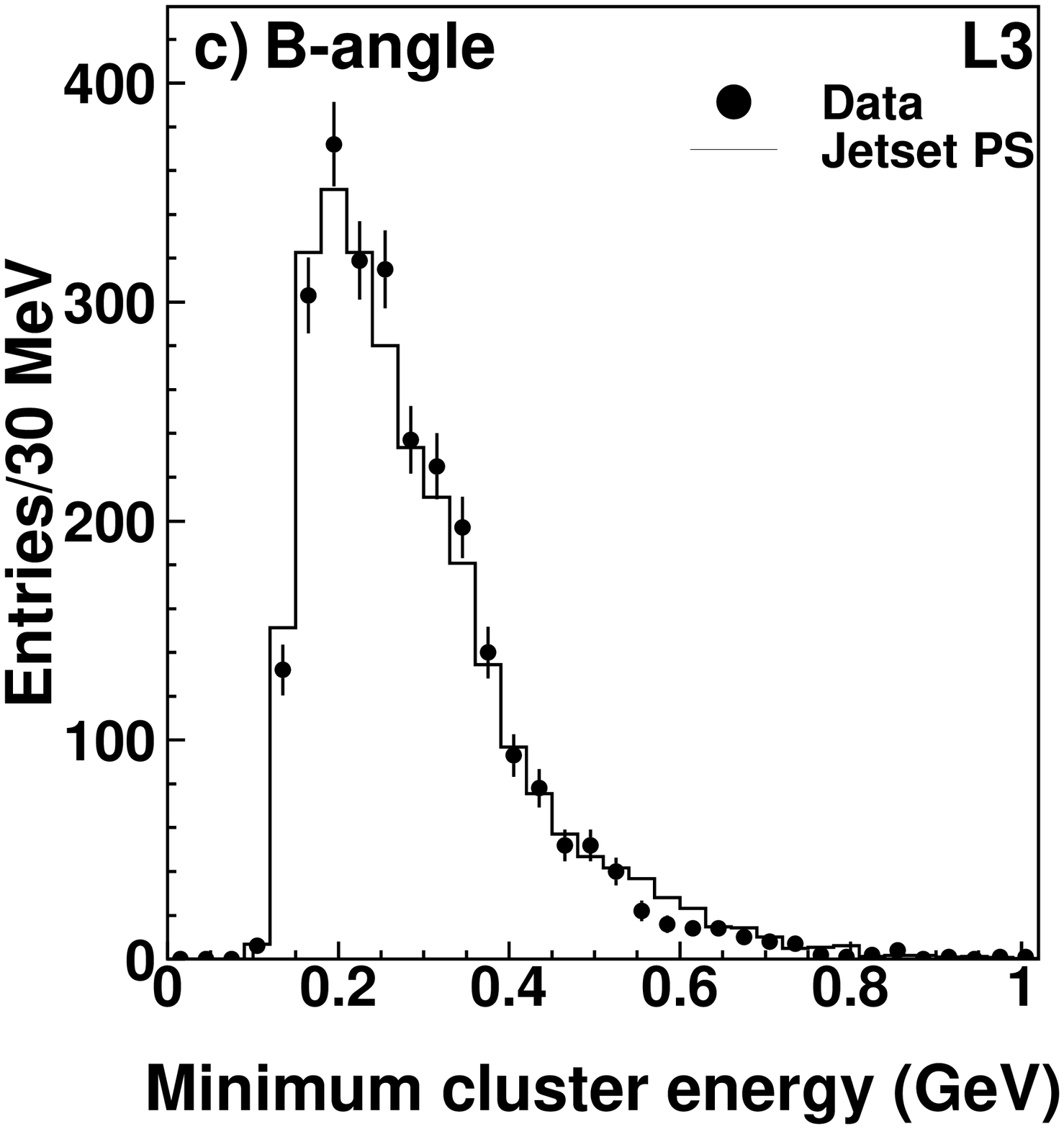}
\includegraphics[width=.48\textwidth]{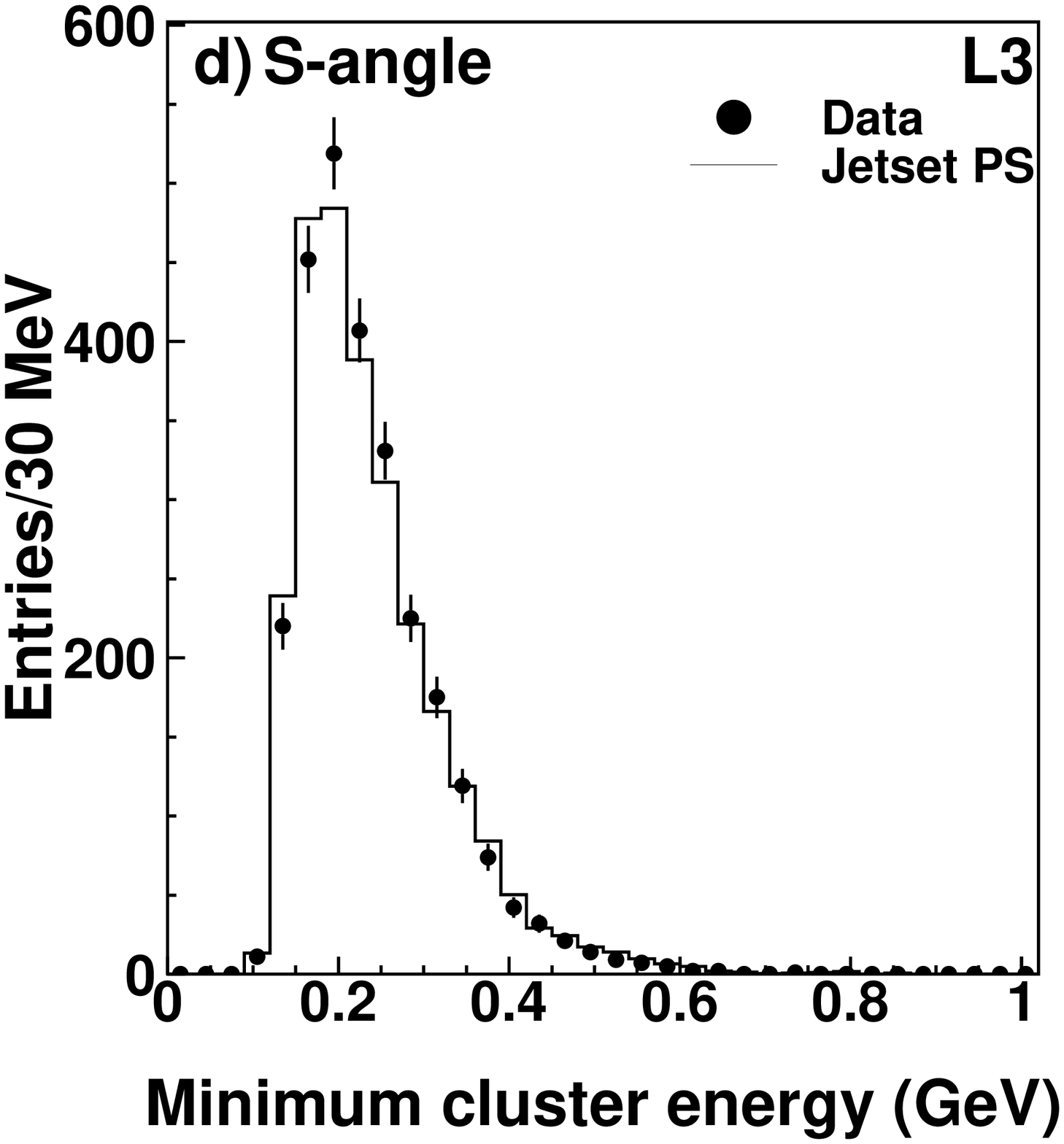}
\caption[Definition of a) minimum angle relative to the gap bisector and b) the
         maximum separation angle between adjacent particles for the case 
         of four particles in the sensitive region of gap 12. Only particles 
         outside the $\pm 15^{\circ}$ cones around the jet axes are 
         considered. Distributions of the minimum energy of clusters used to 
         define c) the bisector angle and d) the maximum separation in 
         selected symmetric three-jet events compared to the \textsc{Jetset}
         PS prediction.]
        {Definition of a) minimum angle relative to the gap bisector and b) the
         maximum separation angle between adjacent particles for the case 
         of four particles in the sensitive region of gap 12. Only particles 
         outside the $\pm 15^{\circ}$ cones around the jet axes are 
         considered. Distributions of the minimum energy of clusters used to 
         define c) the bisector angle and d) the maximum separation in 
         selected symmetric three-jet events compared to the \textsc{Jetset}
         PS prediction.}
\label{fig:gaps}
\end{center}
\end{figure}

 Two gap angle definitions  are used~\cite{sbthesis}, as shown in
  Figures~\ref{fig:gaps}a and \ref{fig:gaps}b: the minimum 
angle, $B_{ij}$, of a particle measured from the bisector in the gap $ij$,
 and the maximum separation angle, $S_{ij}$,  between 
 adjacent particles in the gap.
 In Figures~\ref{fig:gaps}c and \ref{fig:gaps}d the minimum energy of the
 calorimetric clusters used
 to define the bisector angle or the maximum separation angle is compared
 with the \textsc{Jetset} prediction. Good agreement is obtained, showing that 
the contribution of soft particles, to which rapidity gap distributions
  are particularly sensitive, is well simulated. 

 The angular asymmetry in gap 12 from the $B_{ij}$ angles is defined as
$$
\abqq ~=~  \frac{-B_{12}+B_{23}+B_{31}}{B_{12}+B_{23}+B_{31}}~.
$$
 Asymmetries are also defined from the $S_{ij}$ angles as
$$
\asqq ~=~  \frac{-S_{12}+S_{23}+S_{31}}{S_{12}+S_{23}+S_{31}}~.
$$
The other gap asymmetries: $A_{ij}^{B}$,  $A_{ij}^{S}$,  with $ij = 23,31$
  are defined in a similar way.
 
Reduced colour flow and thereby larger separation for \csp in gaps 23, 
31 with respect to gap 12, should thus make $A^B_{12}$ and $A^S_{12}$ peak 
more strongly at positive values for \textsc{Csp} than for \textsc{Cop}. 
The 23 and 31 gap asymmetries of each event are averaged to yield the `qg'
asymmetries shown below. 

The angular asymmetry distributions are corrected for detector effects and 
initial and final state photon radiation
using  bin-by-bin correction factors obtained from events generated with the
\textsc{Jetset}
Parton Shower Monte Carlo program and processed through L3 detector
simulation~\cite{l3-simul}. The bin sizes are chosen sufficiently large
that migration effects are negligible. The correction factors are defined
as the ratio of generated particle-level distributions, considering all 
charged and neutral particles, without energy cuts,
to the same distributions after detector simulation.
 The particle-level distributions take into account the gluon jet 
 identification probability and have a quark
flavour composition  corresponding to $\rs\approx m_{\rm Z}$.
These bin-by-bin correction factors for the 
angular asymmetries typically lie in the range of $\pm 20 \%$.

\begin{figure}[hbtp]
\begin{center}
\includegraphics[width=.48\textwidth]{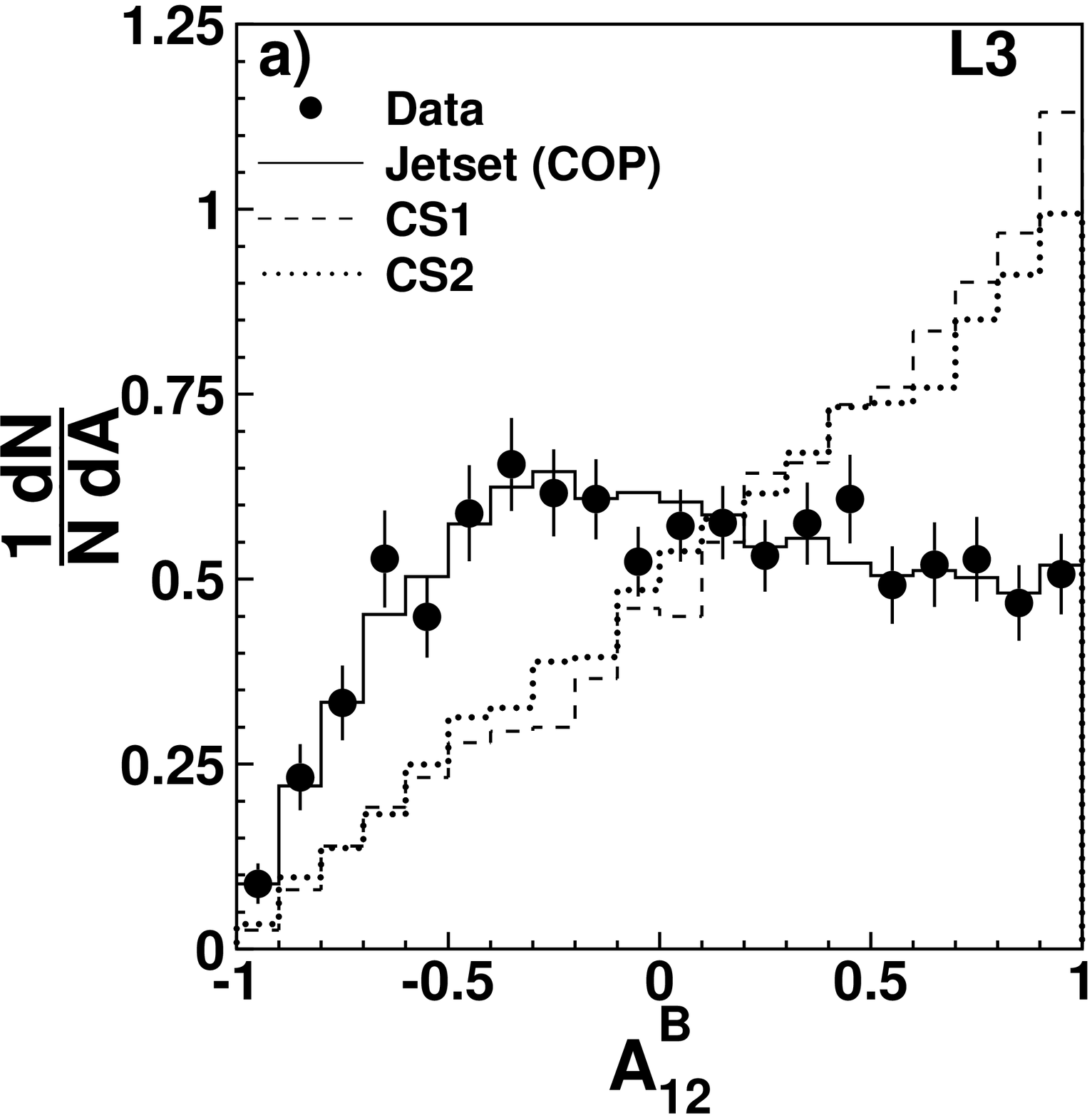}
\includegraphics[width=.48\textwidth]{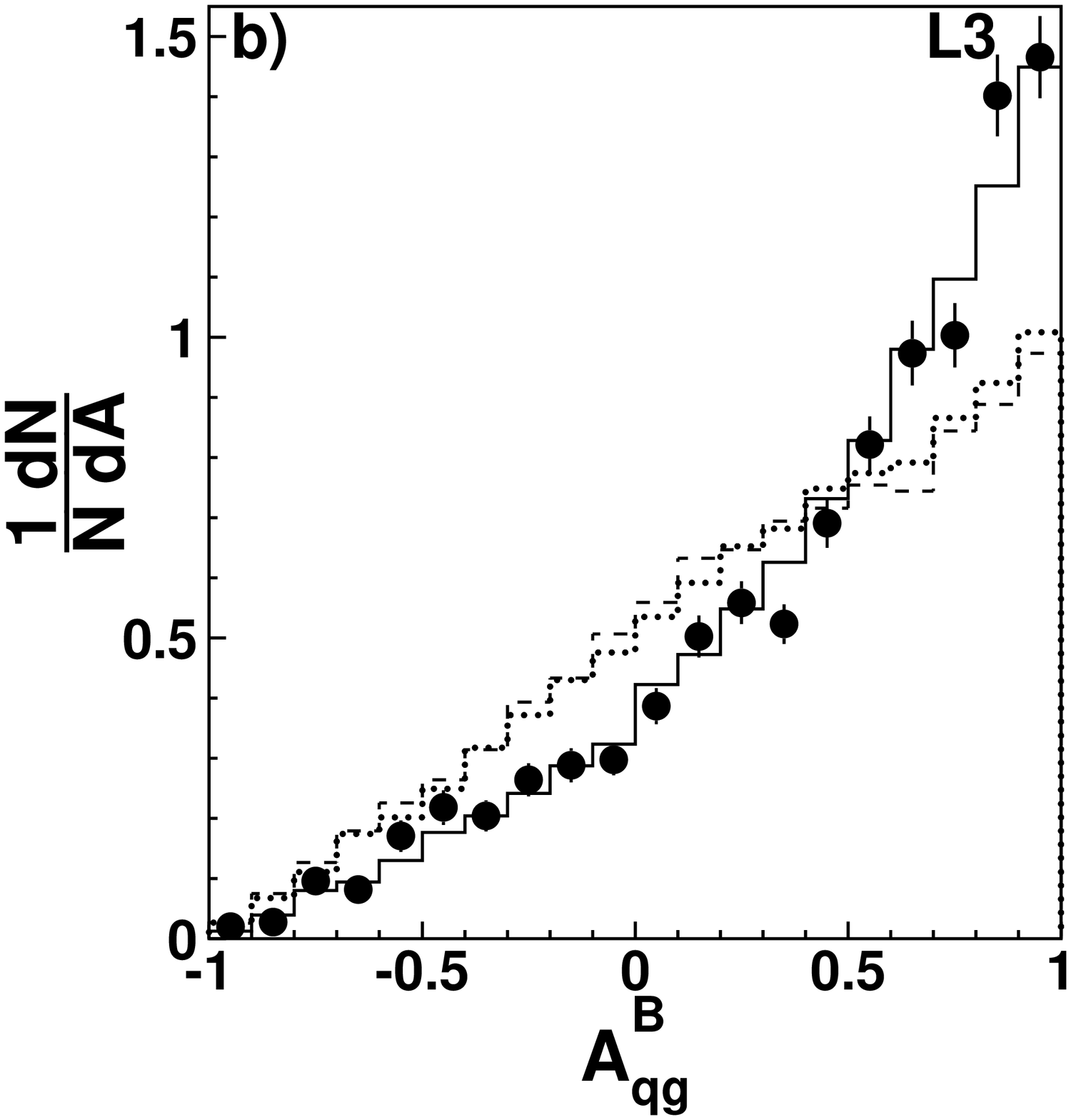}

\vspace*{10mm}
\includegraphics[width=.48\textwidth]{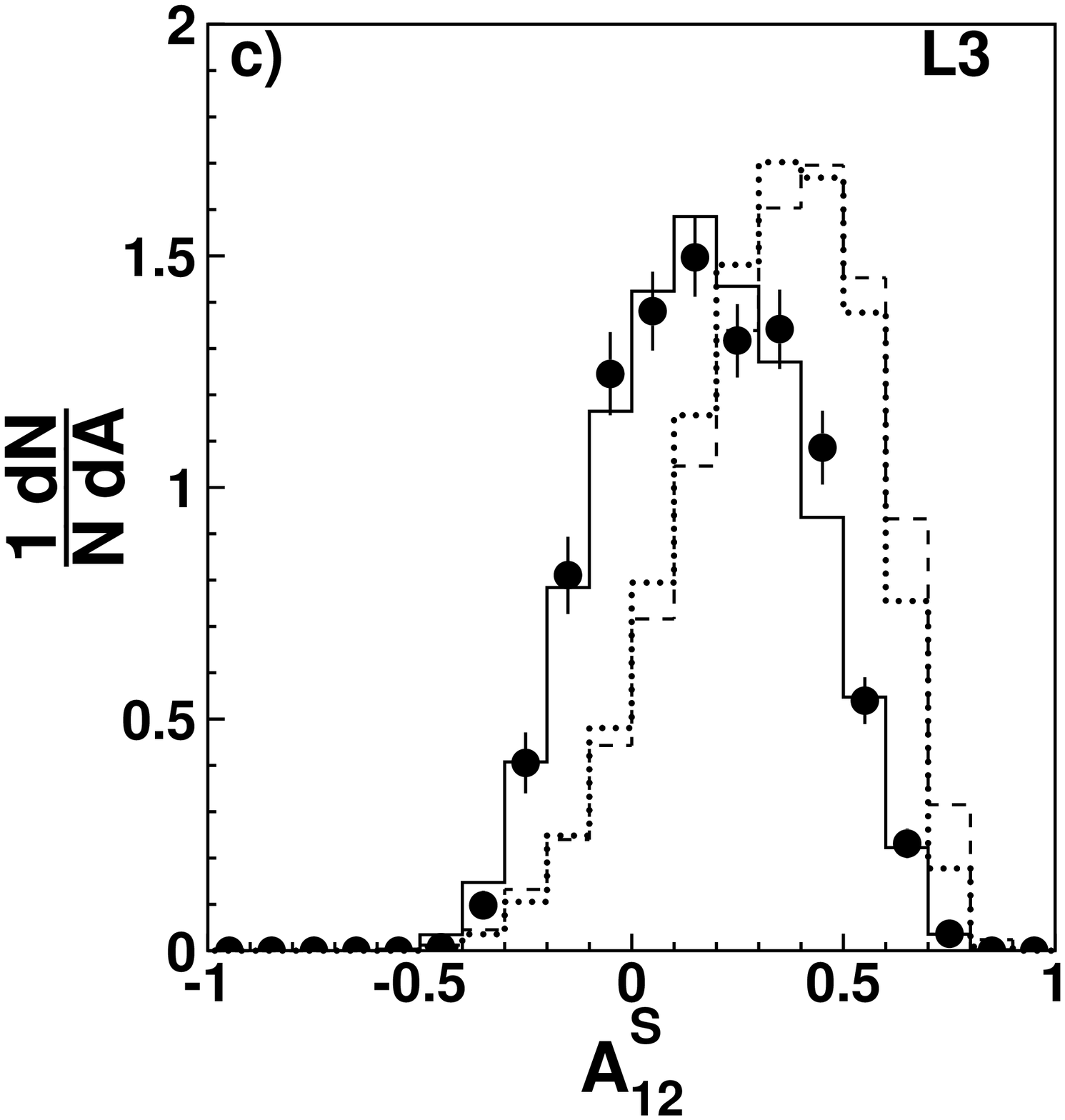}
\includegraphics[width=.48\textwidth]{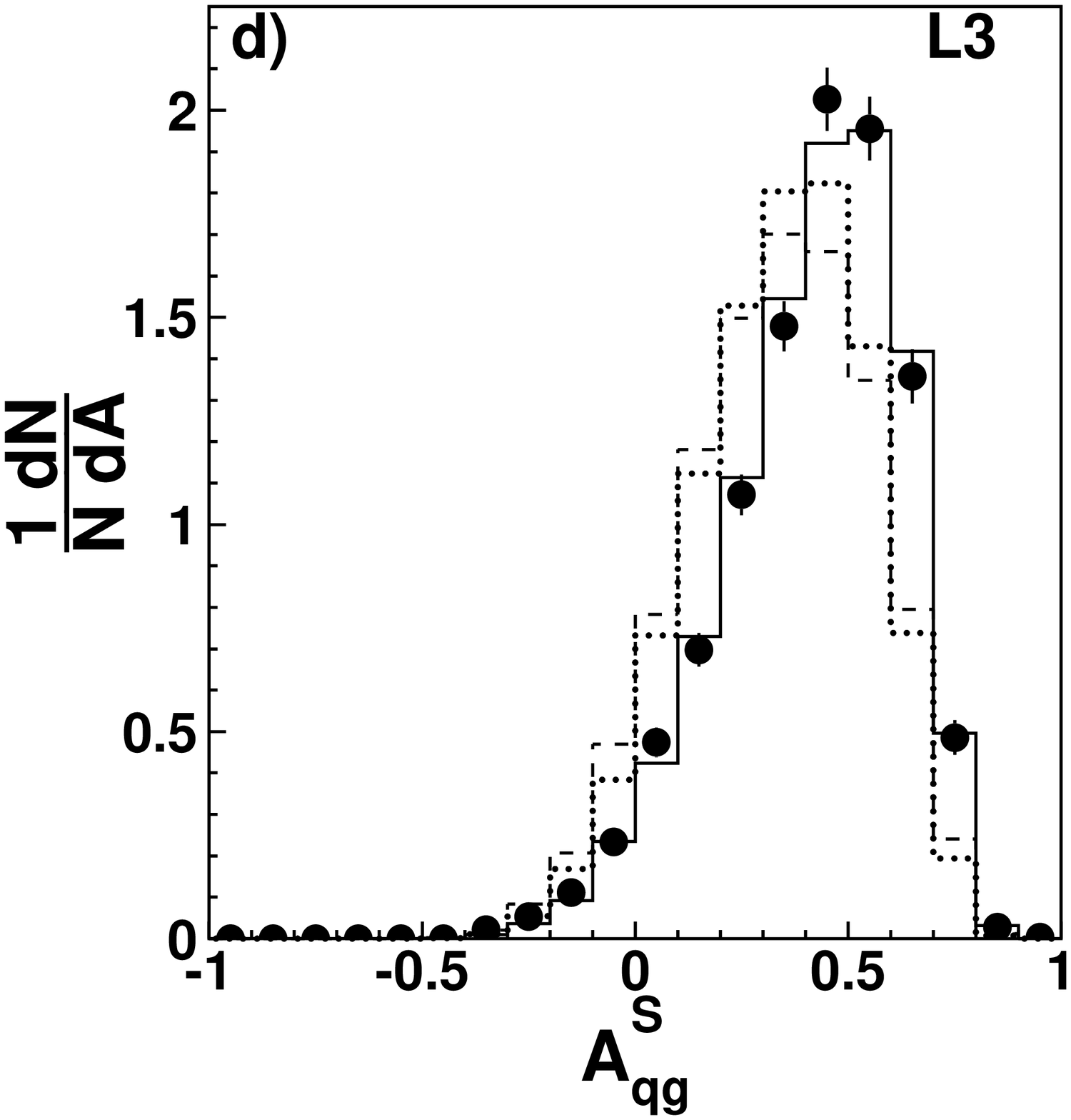}
\caption[a) and b) minimum bisector angle gap asymmetries, and c) and d) 
         maximum separation gap asymmetries for gaps 12 and qg, respectively,
         compared to colour singlet and colour octet models.]
        {a) and b) minimum bisector angle gap asymmetries, and c) and d) 
         maximum separation gap asymmetries for gaps 12 and qg, respectively,
         compared to colour singlet and colour octet models.}
\label{fig:ab}
\end{center}
\end{figure}

\begin{figure}[hbtp]
\begin{center}
\includegraphics[width=.48\textwidth]{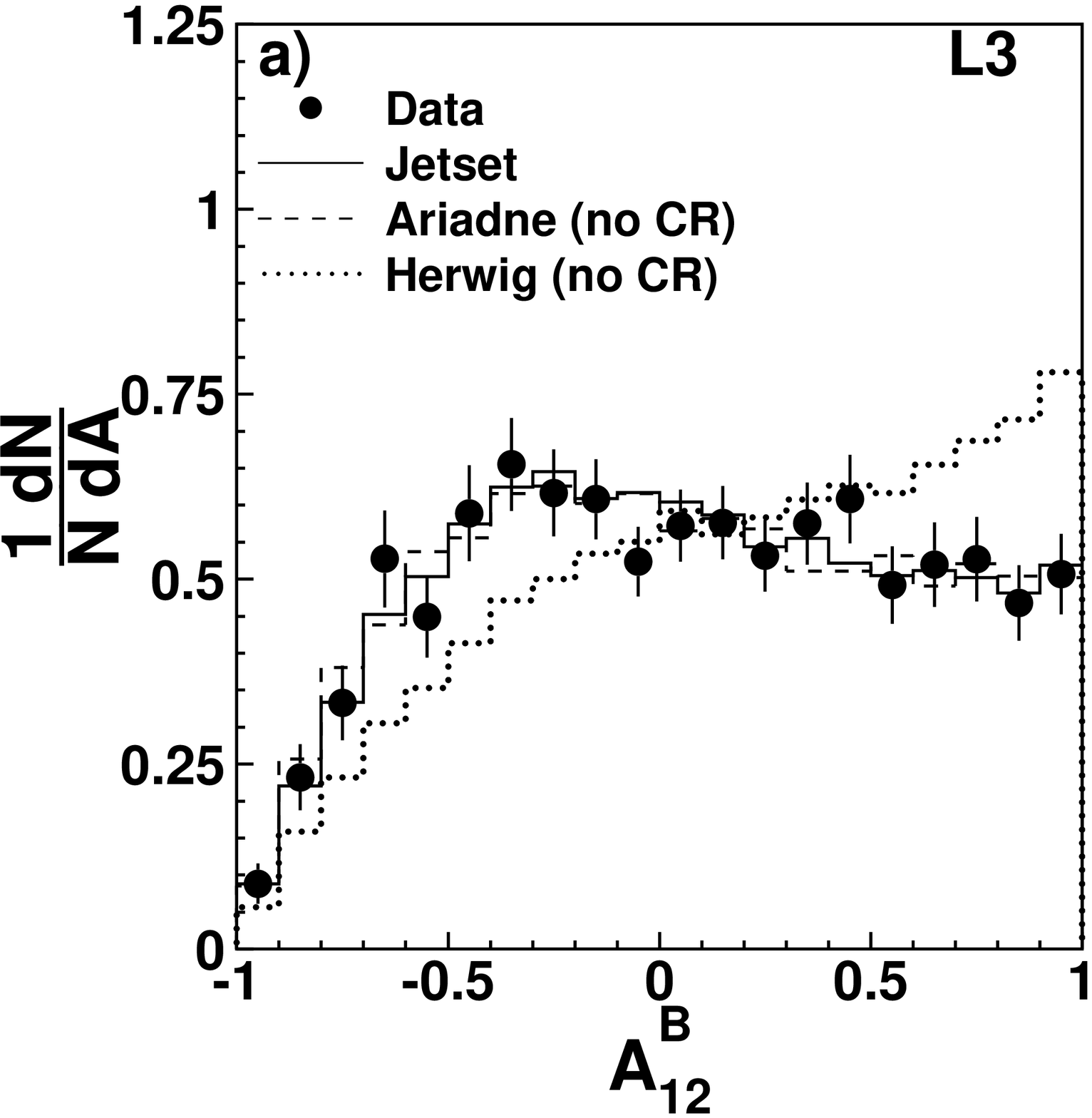}
\includegraphics[width=.48\textwidth]{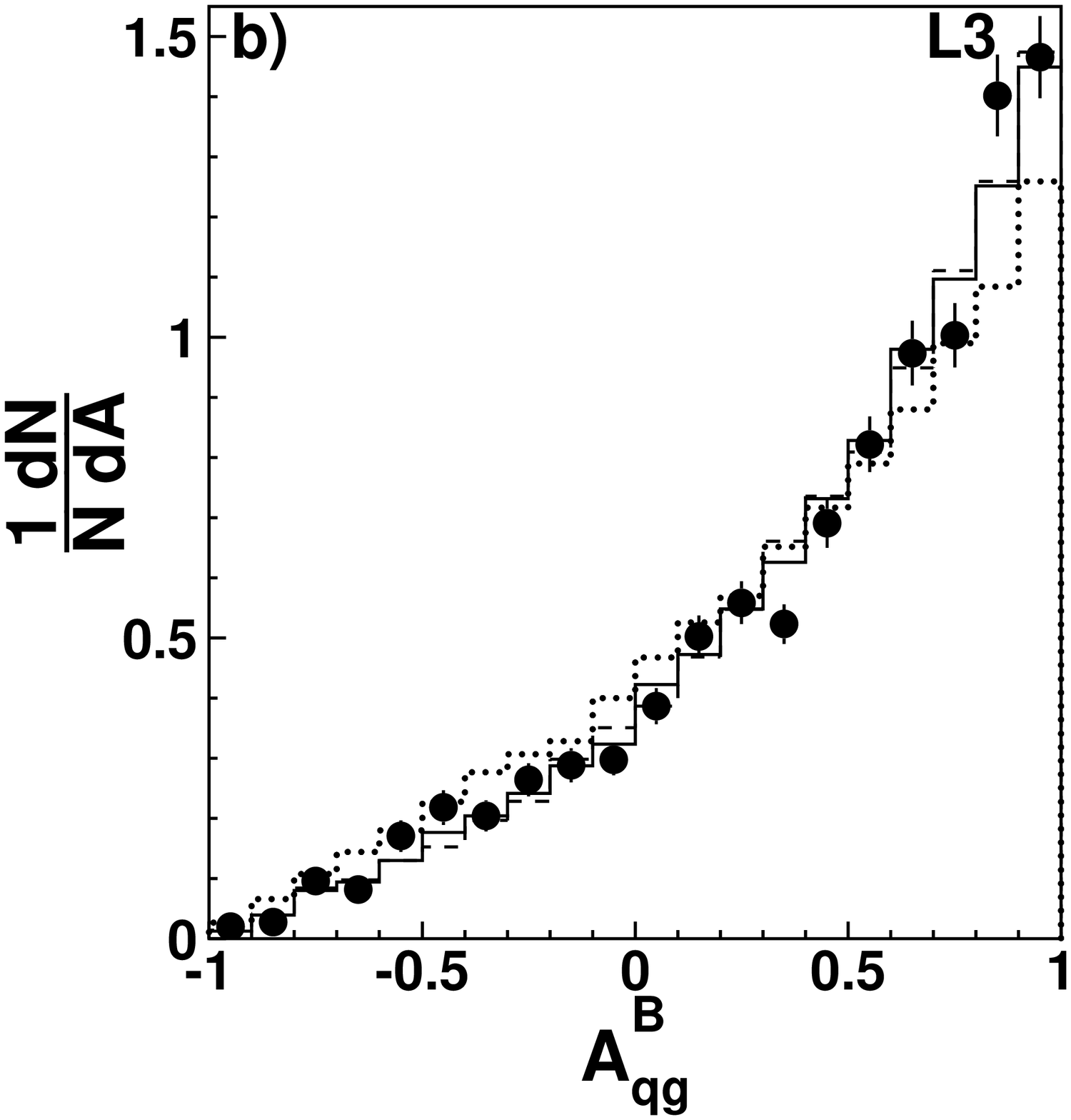}

\vspace*{10mm}
\includegraphics[width=.48\textwidth]{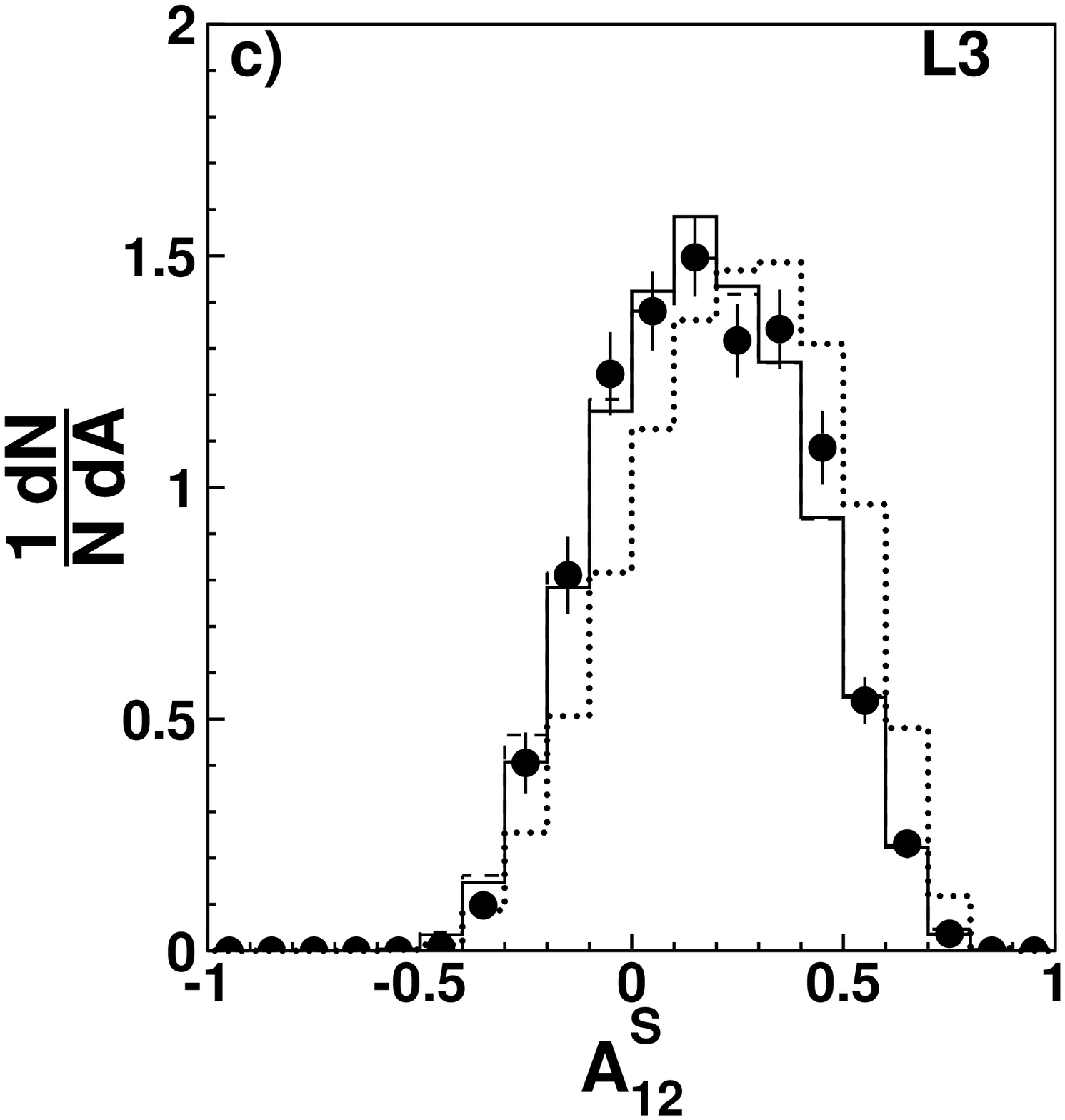}
\includegraphics[width=.48\textwidth]{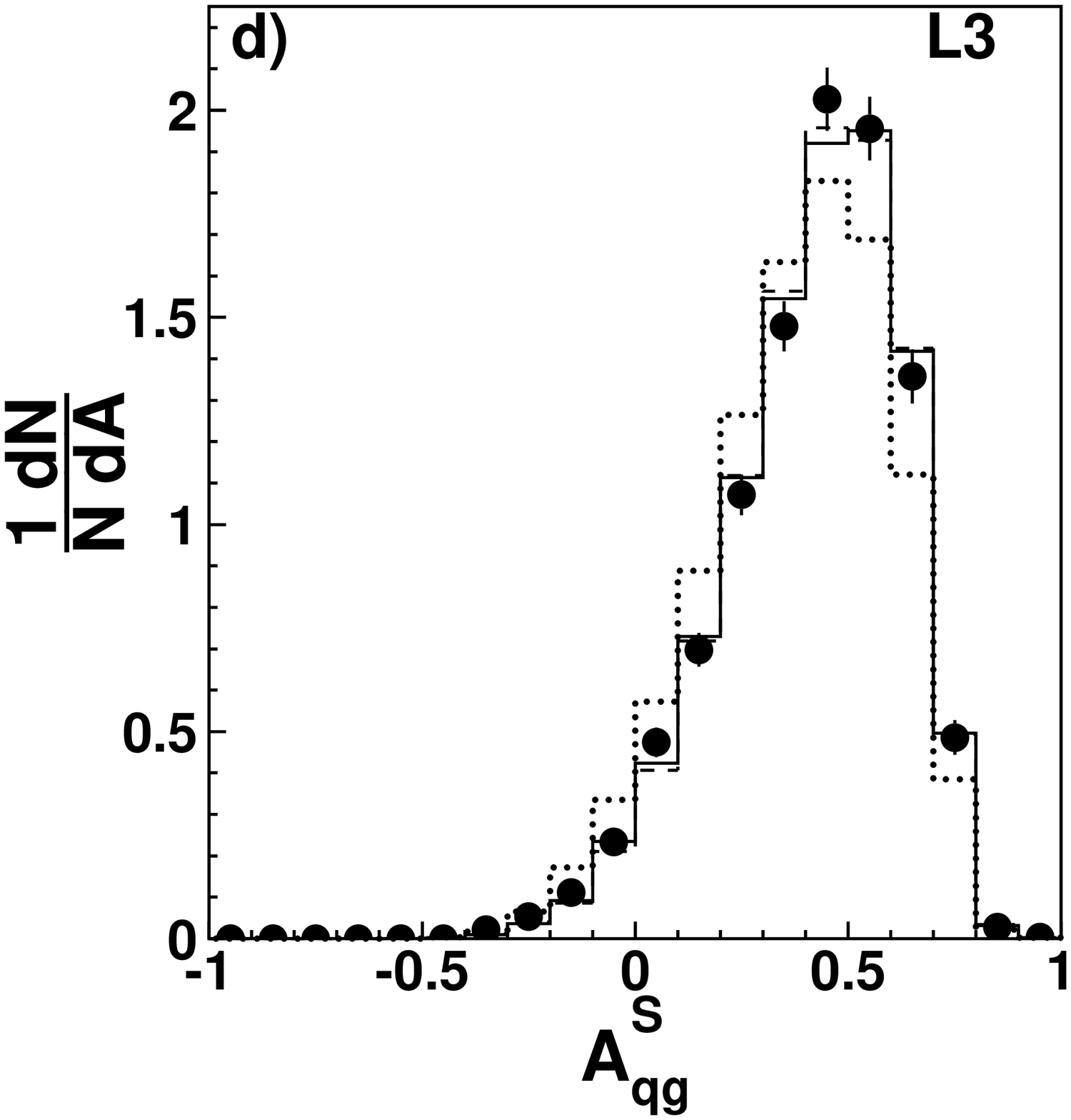}
\caption[a) and b) minimum bisector angle gap asymmetries, and c) and d) 
         maximum separation gap asymmetries for gaps 12 and qg, respectively,
         compared to models without CR effects.]
        {a) and b) minimum bisector angle gap asymmetries, and c) and d) 
         maximum separation gap asymmetries for gaps 12 and qg, respectively,
         compared to models without CR effects.}
\label{fig:ap}
\end{center}
\end{figure}

\begin{figure}[hbtp]
\begin{center}
\includegraphics[width=.48\textwidth]{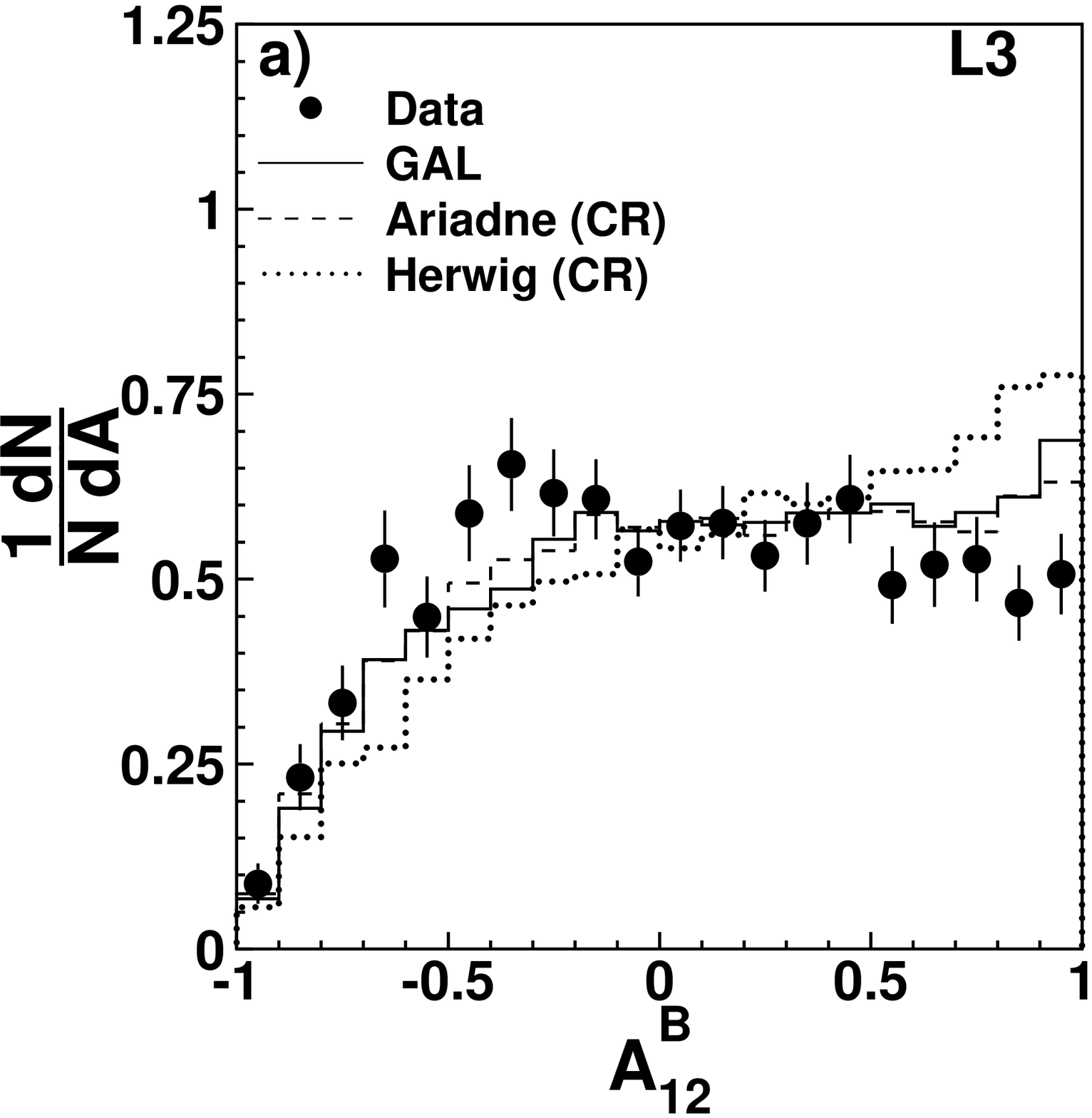}
\includegraphics[width=.48\textwidth]{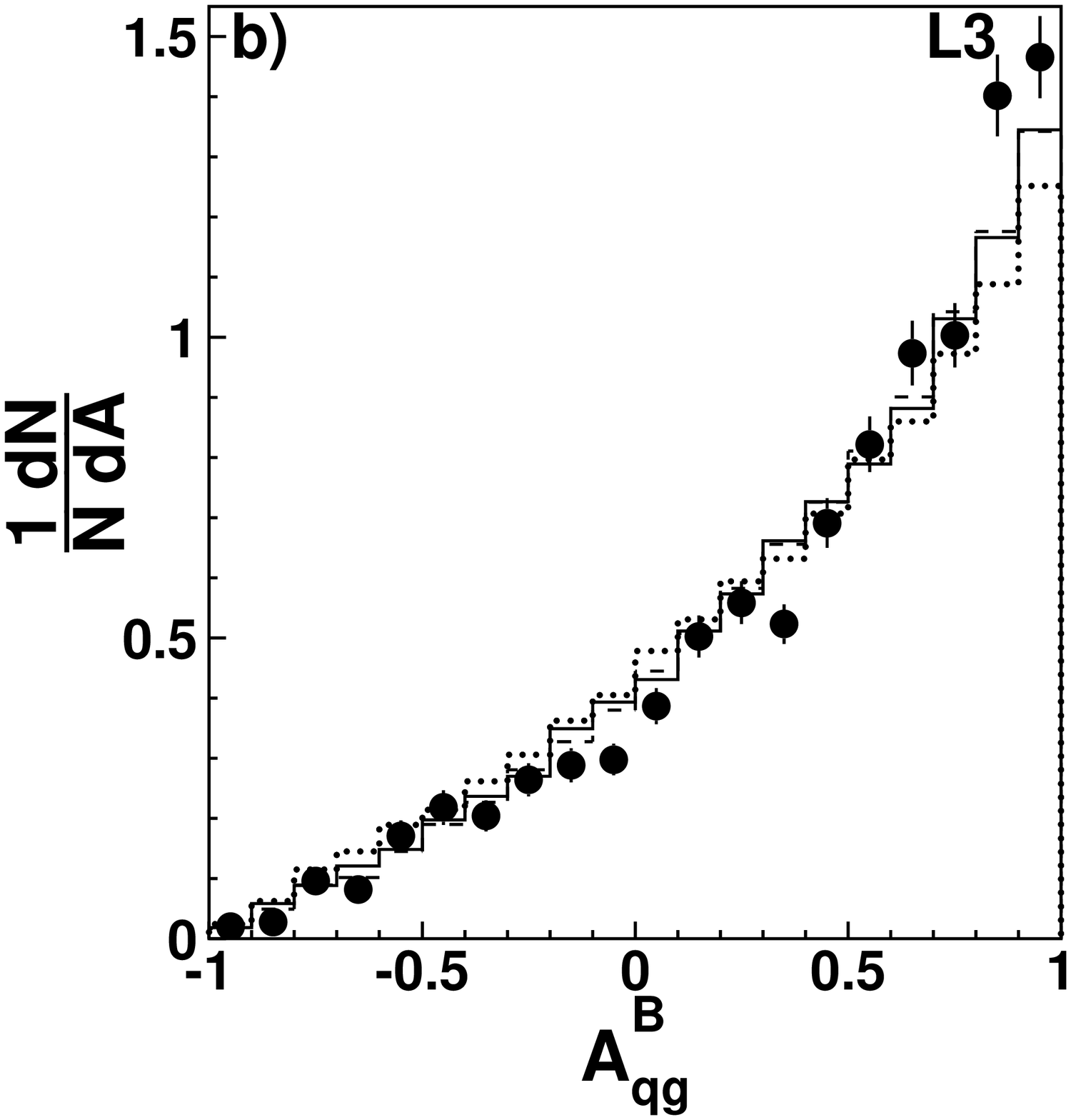}

\vspace*{10mm}
\includegraphics[width=.48\textwidth]{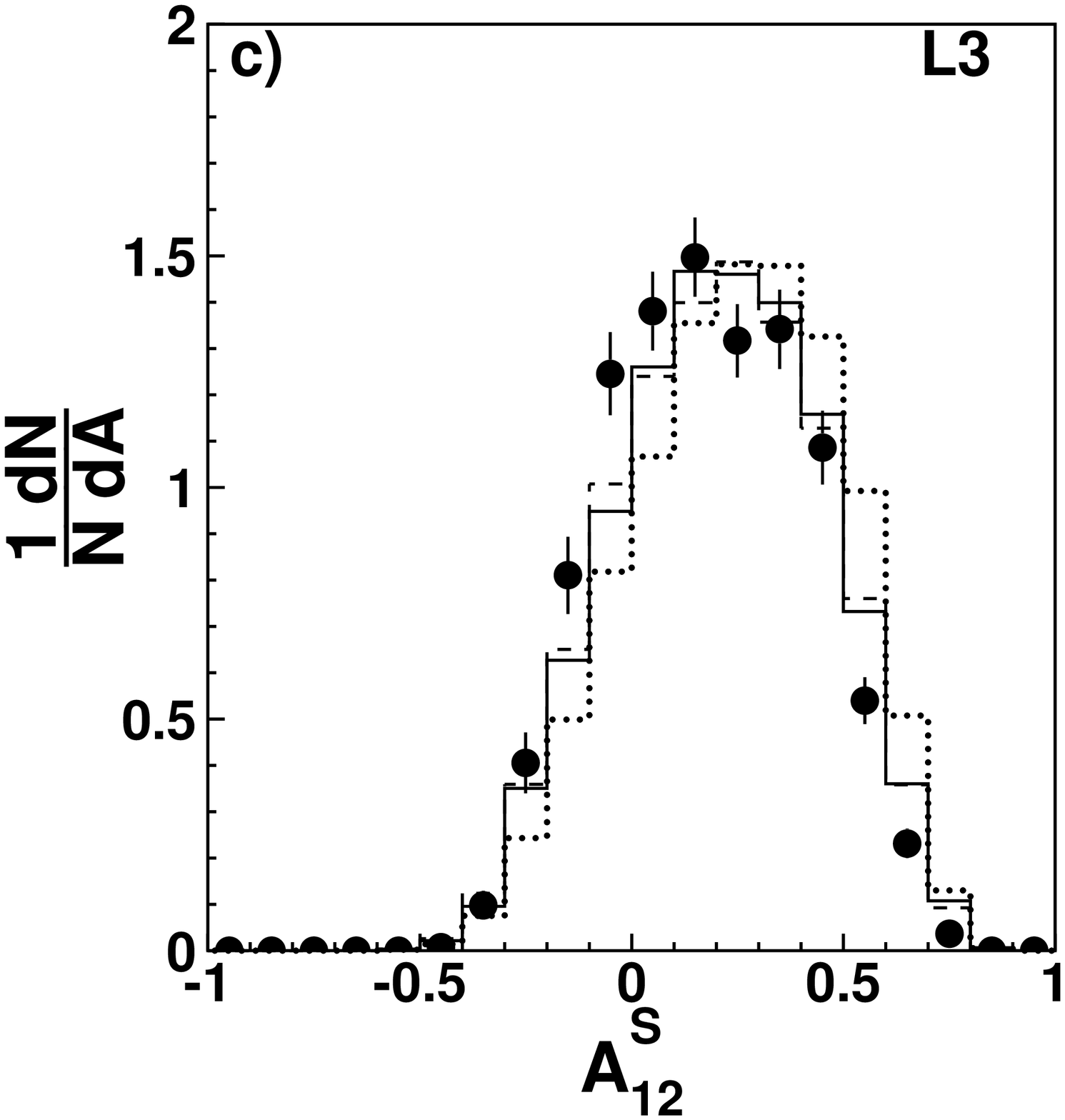}
\includegraphics[width=.48\textwidth]{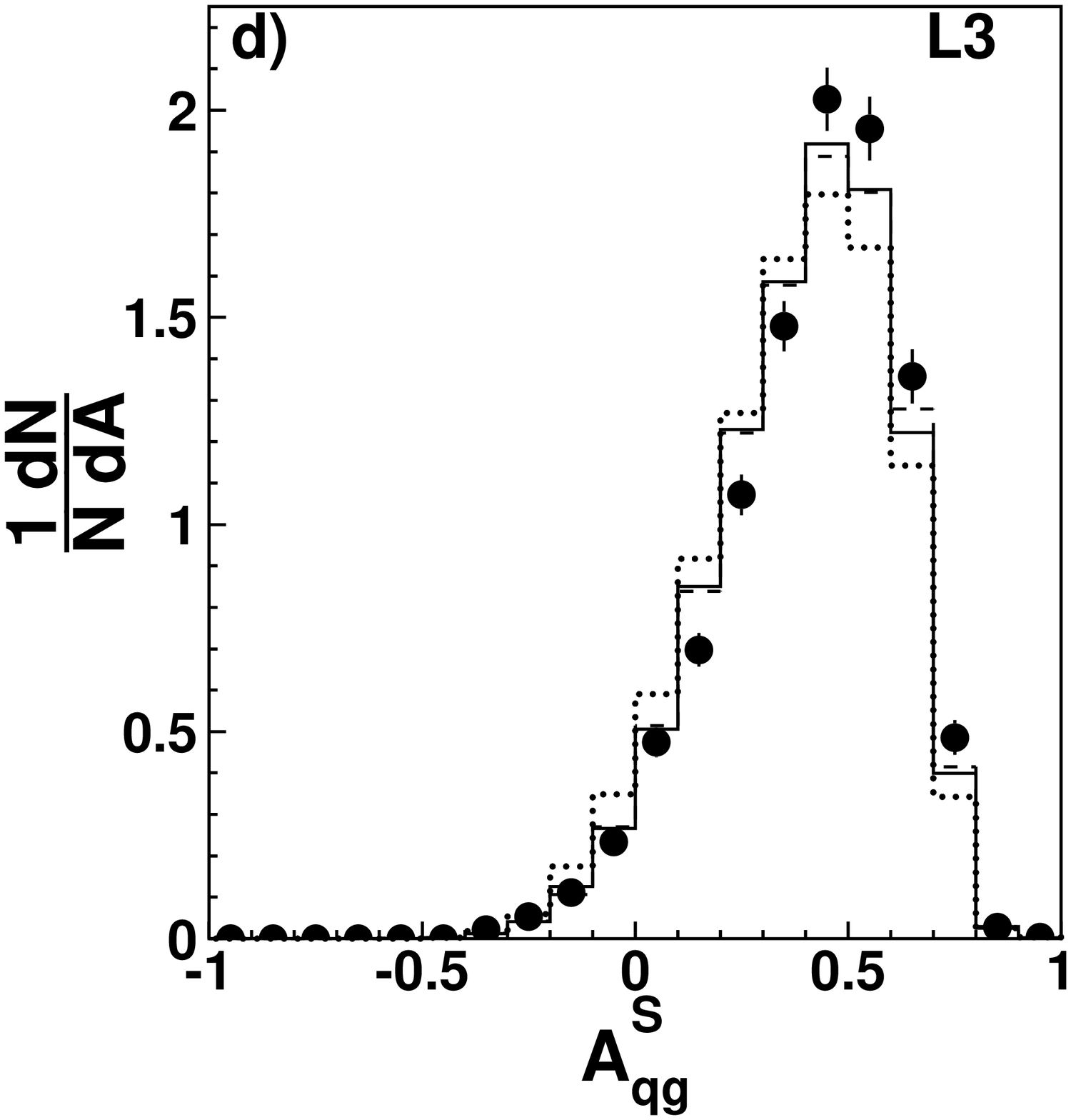}
\caption[a) and b) minimum bisector angle gap asymmetries, and c) and d) 
         maximum separation gap asymmetries for gaps 12 and qg, respectively,
         compared to models with CR effects.]
        {a) and b) minimum bisector angle gap asymmetries, and c) and d) 
         maximum separation gap asymmetries for gaps 12 and qg, respectively,
         compared to models with CR effects.}
\label{fig:as}
\end{center}
\end{figure}

The particle-level angular
asymmetry distributions of the selected symmetric three-jet events, 
normalised to unit area, are compared to different models, in Figures
\ref{fig:ab}, \ref{fig:ap} and \ref{fig:as}. 

Fractional 
bin-by-bin systematic uncertainties are estimated by repeating the analysis
using clusters obtained by combining calorimetric clusters with
tracks, as used, for instance, in Reference~\citen{DDWCR}. A variation
between 2\% and 5\% is observed. Furthermore, the cuts on the b-tag
discriminant are changed so that the gluon purity varies by $\pm 10$\%, 
which results in systematic uncertainties between 3\% and
8\%. Finally, the residual 2.5\% difference between data and Monte
Carlo discussed above is included.  These uncertainties are added in 
quadrature and are summarised in Table \ref{tab:syst}. 
The systematic 
uncertainty due to a  change in the jet cone angle cut from $\pm 15^{\circ}$ 
to $\pm 20^{\circ}$ is found to be negligible. 

\begin{table}[hbtp]
\begin{center}
\begin{tabular}{|c|c|c|c|c|}\hline
      & \multicolumn{4}{c|}{Systematic uncertainties (\%) }\\ \cline{2-5}
  Variable&  Detector & B-tag & Monte Carlo &  Total \\ \hline
\rule{0pt}{12pt}
A$^{\B}_{12}$   & $5.2$ & $4.8$  & $2.5$ & $\phantom{0} 7.5$  \\
\rule{0pt}{12pt}
A$^{\B}_{\qg}$  & $5.9$ & $3.2$  & $2.5$ & $\phantom{0} 7.1$  \\
\rule{0pt}{12pt}
A$^{\S}_{12}$   & $6.6$ & $8.1$  & $2.5$ & $10.8$  \\
\rule{0pt}{12pt}
A$^{\S}_{\qg}$  & $2.8$ & $4.0$  & $2.5$ & $\phantom{0} 5.5$  \\ \hline
\end{tabular}
\caption[]{Systematic uncertainties on the measurements of the asymmetry 
           variables}
\label{tab:syst}
\end{center}
\end{table}
            
\section{Limits on Colour Singlet Production}

As shown in Figure \ref{fig:ab}, the data are in good agreement with the \cop 
model. The high discrimination power of the angular asymmetries between the 
\cop and \csp models is also evident. A quantitative comparison is given in 
Table \ref{tab:chi2no1}. The \cop model is in good agreement with the data.
The \csp models are clearly excluded.
As a cross-check, the analysis is repeated using the {\sc Durham}
$k_\perp$ algorithm~\cite{kt} with $y_{cut}=0.01$ and $y_{cut}=0.02$
instead of the {\sc Jade } algorithm, effectively defining an
independent set of asymmetries. No significant changes are observed.

\begin{table}[htbp]
\begin{center}
\begin{tabular}{|c|c|c|c|c|} \hline
  &  \multicolumn{3}{c|}{$\chi^{2}$ for} &  \\ \cline{2-4}
Variable & \cop & \textsc{Cs1} & \textsc{Cs2} & d.o.f. \\ \hline
\rule{0pt}{12pt}
A$^{\B}_{12}$  & $\phantom{0} 6.4~(0.99)$  & $\phantom{0} 356$  &  262 & 19  \\
\rule{0pt}{12pt}
A$^{\B}_{\qg}$ & $15.9~(0.60)$  &  $\phantom{0} 238$  &   189 & 18 \\
\rule{0pt}{12pt}
A$^{\S}_{12}$  & $\phantom{0}  4.8~(0.94)$  &  1081 &  524 & 11  \\
\rule{0pt}{12pt}
A$^{\S}_{\qg}$ & $\phantom{0}6.7~(0.88)$  &  $\phantom{0} 334$  &  266 & 12  \\
\hline\end{tabular}
\caption[]{Values of $\chi^{2}$  obtained from the comparison of the data
           distributions to colour octet and colour singlet models.
           For \cop the corresponding confidence levels are given in 
           parentheses. For the \csp models all confidence levels are less 
           than $ 10^{-30}$. The $\chi^{2}$ values include systematic 
           uncertainties.}
\label{tab:chi2no1}
\end{center}
\end{table}

The asymmetry distributions are fitted to a combination of \cop  and  \csp 
contributions. This is done by minimising a $\chi^{2}$  function defined as
\begin{eqnarray*}
\chi^2 (r) & = & \sum_{i}\frac{[ f_{data}^i - r f_{\csp}^i - 
                (1-r) f_{\cop}^i ]^2}{ (\sigma_{stat}^i)^2 + 
                (\sigma_{syst}^i)^2}
\end{eqnarray*}
where $f^i$ is the content of the $i$th bin of the distribution, $r$ is the 
fraction of the \csp component and contributions from both statistical, 
$\sigma_{stat}^i$,  and systematic, $\sigma_{syst}^i$, uncertainties are 
included. 

Good fits are obtained for all the asymmetry distributions.
For the variable A$^{\S}_{12}$, which Monte Carlo studies show to be the 
most sensitive one, the fit gives $r$ = 0.015 $\pm$ 0.024 (stat.) $\pm$ 
0.018 (syst.) with $\chi^{2}$/d.o.f. = 4.5/11 for the \textsc{Cs1} model
$r$ = 0.025 $\pm$ 0.031 (stat.) $\pm$ 0.029 (syst.) with $\chi^{2}$/d.o.f.
= 4.4/11 for the \textsc{Cs2} model. All the fits 
give a fraction of events due to \csp consistent with zero.
The fits to the distributions are then used to obtain a 95\% confidence 
level (CL) upper  bound on the fraction of \csp events. 
The asymmetry variables $A^{\B}_{12}$ and
$A^{\B}_{\qg}$ are independent, as are $A^{\S}_{12}$ and $A^{\S}_{\qg}$.
These pairs of variables are thus combined in the fits. Upper 
bounds of 6.7\% and 10.2\% for the \textsc{Cs1} and \textsc{Cs2} models, 
respectively, are found using $A^{\S}_{12}$ and $A^{\S}_{\qg}$. Using
$A^{\B}_{12}$ and $A^{\B}_{\qg}$  yields slightly weaker limits.

\section{Limits on Colour Reconnection Effects}

The particle-level angular asymmetry distributions are compared to the
predictions of several different Monte Carlo models in Figures
\ref{fig:ap} and \ref{fig:as}. These include the `no CR' models
\textsc{Jetset}, \textsc{Ariadne} and \textsc{Herwig} as well as the
\textsc{Gal} model and the CR versions of \textsc{Ariadne} and
\textsc{Herwig}.  The $\chi^2$ confidence levels (CLs) given by the
comparison of these models to the data are presented in Table
\ref{tab:chi2no2}. Both the default \textsc{Gal} and \textsc{Ariadne}
CR models are excluded by the $A^{\S}_{12}$ distribution, with CLs of
$\simeq 10^{-8}$ and $\simeq 10^{-6}$ respectively. The \textsc{Gal}
model is also excluded by the $A^{\B}_{\qg}$ distribution which gives
CL $\simeq 10^{-4}$. The same distribution has a low CL of $ \simeq
10^{-2}$ for the \textsc{Ariadne} CR model. However \textsc{Ariadne},
without CR, gives a satisfactory description of all of the
distributions. Both versions of \textsc{Herwig} are completely
excluded, with a best CL of $\simeq 10^{-8}$ for no CR and of $\simeq
10^{-9}$ for CR, among all of the asymmetry distributions considered,
suggesting that it cannot be used, with confidence, to simulate the
soft hadronisation effects that are important for CR
studies\footnote{This conclusion does not depend on the size of the
systematic uncertainties which, even doubled, would still give
\textsc{Herwig} CLs of $\simeq 10^{-7}$ for the $A^{\S}_{12}$ and
$A^{\S}_{\qg}$ asymmetries.}.  Consistent results are obtained by
repeating the analysis using the {\sc Durham} $k_\perp$
algorithm~\cite{kt} with $y_{cut}=0.01$ and $y_{cut}=0.02$ instead of
the {\sc Jade } algorithm.

\begin{table}[htbp]
\begin{center}
{\small\begin{tabular}{|c|c|c|c|c|c|c|} \hline
   & \multicolumn{3}{c|}{ No CR} & \multicolumn{3}{c|}{CR}  \\ \cline{2-7}
Variable & \textsc{Jetset} & \textsc{Ariadne} & \textsc{Herwig} &
 \textsc{Gal} &  \textsc{Ariadne} & \textsc{Herwig} \\ \hline
\rule{0pt}{12pt}
$A^{\B}_{12}$  & $ 0.99$  & $ 0.93$ & $10^{-9}\phantom{0}$ & $0.04$ & $0.27$ &
 $ 10^{-11}$ \\
\rule{0pt}{12pt}
$A^{\B}_{\qg}$ &  $0.60$ & $0.13$ & $10^{-8}\phantom{0}$ &  $10^{-4}$ &
 $ 0.02 $ & $10^{-8}\phantom{0}$ \\
\rule{0pt}{12pt}
$A^{\S}_{12}$  & $ 0.94 $ & $ 0.80 $ & $ 10^{-24}$ & $ 10^{-8}$ &
 $ 10^{-6}$  & $ 10^{-30}$ \\
\rule{0pt}{12pt}
$A^{\S}_{\qg}$ & $ 0.88 $  & $ 0.78 $ & $10^{-8}\phantom{0}$ & $ 0.03 $ &
  $ 0.07 $  & $ 10^{-11}$ \\ 
\hline\end{tabular}}
\caption[]{$\chi^2$ confidence levels obtained from the comparison of the data
           distributions to different models with and without CR. Low 
           confidence levels are rounded to the nearest order of magnitude.}
\label{tab:chi2no2}
\end{center}
\end{table}

Fits are performed to the asymmetry distributions to obtain the best 
value of $R_0$ by interpolating the Monte Carlo distributions with 
different values of $R_0$. Good fits are obtained, in all cases, with values 
of $R_0$ consistent with zero. Further fits are then performed to 
obtain an upper limit on $R_0$. Combining the pair of variables $A^{\B}_{12}$ 
and $A^{\B}_{\qg}$ or $A^{\S}_{12}$ and $A^{\S}_{\qg}$, a 95\% CL upper limit 
for $R_0$ of 0.024 is obtained.

\section{Summary and Conclusions}
 
New observables based on angular separations of particles in the inter-jet
regions of symmetric three-jet events are introduced and are found to be
very sensitive to \csp and to CR effects.

Upper limits at 95\% CL on \csp according to the \textsc{Cs1} and \textsc{Cs2}
models of 6.7\% and  10.2\%, respectively, are obtained. Since the fraction of 
\csp expected on the basis of the \textsc{Tevatron} measurements is only 
5--10\%, after allowing for the effect 
of gap survival probability, the present analysis is not sufficiently 
sensitive to confirm or exclude a similar effect in hadronic Z decays.

The \textsc{Gal} model, with the default CR probability,
and the \textsc{Ariadne} CR model are unable to describe the data. Both the
no CR and CR versions of \textsc{Herwig} are completely excluded by the data.
However, a good description is provided both by \textsc{Jetset} and the no CR
version of \textsc{Ariadne}. This suggests that the angular asymmetries are 
also very sensitive to the non-perturbative hadronisation model used. Both
\textsc{Jetset} and \textsc{Ariadne} have similar, string-like, hadronisation
models, whereas \textsc{Herwig} uses cluster fragmentation.

The results presented in this Letter provide important information
concerning the systematic uncertainty on the W mass resulting from
CR effects as estimated by the \textsc{Gal}, \textsc{Ariadne}
and  \textsc{Herwig} Monte Carlo models. For the default value of the
\textsc{Gal} CR parameter, $R_0 = 0.1$, the W mass measured from decays of 
W pairs into four jets is estimated~\cite{rath} to be shifted by
about 65 MeV. The 0.024 95\% CL upper limit obtained in this analysis
implies a mass shift of only a few MeV. Since the default CR models in 
\textsc{Ariadne} and \textsc{Herwig} are unable to correctly describe the 
Z-decay data, it is difficult to have confidence in their use to describe
CR effects in W-pair production.

\newpage
\section*{Author List}
\typeout{   }     
\typeout{Using author list for paper 261 -  }
\typeout{$Modified: Jul 15 2001 by smele $}
\typeout{!!!!  This should only be used with document option a4p!!!!}
\typeout{   }
%
%
%
%
%
%

\newcount\tutecount  \tutecount=0
\def\tutenum#1{\global\advance\tutecount by 1 \xdef#1{\the\tutecount}}
\def\tute#1{$^{#1}$}
\tutenum\aachen            
\tutenum\nikhef            
\tutenum\mich              
\tutenum\lapp              
\tutenum\basel             
\tutenum\lsu               
\tutenum\beijing           
\tutenum\bologna           
\tutenum\tata              
\tutenum\ne                
\tutenum\bucharest         
\tutenum\budapest          
\tutenum\mit               
\tutenum\panjab            
\tutenum\debrecen          
\tutenum\dublin            
\tutenum\florence          
\tutenum\cern              
\tutenum\wl                
\tutenum\geneva            
\tutenum\hefei             
\tutenum\lausanne          
\tutenum\lyon              
\tutenum\madrid            
\tutenum\florida           
\tutenum\milan             
\tutenum\moscow            
\tutenum\naples            
\tutenum\cyprus            
\tutenum\nymegen           
\tutenum\caltech           
\tutenum\perugia           
\tutenum\peters            
\tutenum\cmu               
\tutenum\potenza           
\tutenum\prince            
\tutenum\riverside         
\tutenum\rome              
\tutenum\salerno           
\tutenum\ucsd              
\tutenum\sofia             
\tutenum\korea             
\tutenum\purdue            
\tutenum\psinst            
\tutenum\zeuthen           
\tutenum\eth               
\tutenum\hamburg           
\tutenum\taiwan            
\tutenum\tsinghua          

{
\parskip=0pt
\noindent
{\bf The L3 Collaboration:}
\ifx\selectfont\undefined
 \baselineskip=10.8pt
 \baselineskip\baselinestretch\baselineskip
 \normalbaselineskip\baselineskip
 \ixpt
\else
 \fontsize{9}{10.8pt}\selectfont
\fi
\medskip
\tolerance=10000
\hbadness=5000
\raggedright
\hsize=162truemm\hoffset=0mm
\def\r{\rlap,}
\noindent

P.Achard\r\tute\geneva\ 
O.Adriani\r\tute{\florence}\ 
M.Aguilar-Benitez\r\tute\madrid\ 
J.Alcaraz\r\tute{\madrid}\ 
G.Alemanni\r\tute\lausanne\
J.Allaby\r\tute\cern\
A.Aloisio\r\tute\naples\ 
M.G.Alviggi\r\tute\naples\
H.Anderhub\r\tute\eth\ 
V.P.Andreev\r\tute{\lsu,\peters}\
% ...
\newpage
\begin{list}{A}{\itemsep=0pt plus 0pt minus 0pt\parsep=0pt plus 0pt minus 0pt
                \topsep=0pt plus 0pt minus 0pt}
\item[\aachen]
 III. Physikalisches Institut, RWTH, D-52056 Aachen, Germany$^{\S}$
\item[\nikhef] National Institute for High Energy Physics, NIKHEF, 
     and University of Amsterdam, NL-1009 DB Amsterdam, The Netherlands
\item[\mich] University of Michigan, Ann Arbor, MI 48109, USA
\item[\lapp] Laboratoire d'Annecy-le-Vieux de Physique des Particules, 
     LAPP,IN2P3-CNRS, BP 110, F-74941 Annecy-le-Vieux CEDEX, France
\item[\basel] Institute of Physics, University of Basel, CH-4056 Basel,
     Switzerland
\item[\lsu] Louisiana State University, Baton Rouge, LA 70803, USA
\item[\beijing] Institute of High Energy Physics, IHEP, 
  100039 Beijing, China$^{\triangle}$ 
\item[\bologna] University of Bologna and INFN-Sezione di Bologna, 
     I-40126 Bologna, Italy
\item[\tata] Tata Institute of Fundamental Research, Mumbai (Bombay) 400 005, India
\item[\ne] Northeastern University, Boston, MA 02115, USA
\item[\bucharest] Institute of Atomic Physics and University of Bucharest,
     R-76900 Bucharest, Romania
\item[\budapest] Central Research Institute for Physics of the 
     Hungarian Academy of Sciences, H-1525 Budapest 114, Hungary$^{\ddag}$
\item[\mit] Massachusetts Institute of Technology, Cambridge, MA 02139, USA
\item[\panjab] Panjab University, Chandigarh 160 014, India.
\item[\debrecen] KLTE-ATOMKI, H-4010 Debrecen, Hungary$^\P$
\item[\dublin] Department of Experimental Physics,
  University College Dublin, Belfield, Dublin 4, Ireland
\item[\florence] INFN Sezione di Firenze and University of Florence, 
     I-50125 Florence, Italy
\item[\cern] European Laboratory for Particle Physics, CERN, 
     CH-1211 Geneva 23, Switzerland
\item[\wl] World Laboratory, FBLJA  Project, CH-1211 Geneva 23, Switzerland
\item[\geneva] University of Geneva, CH-1211 Geneva 4, Switzerland
\item[\hefei] Chinese University of Science and Technology, USTC,
      Hefei, Anhui 230 029, China$^{\triangle}$
\item[\lausanne] University of Lausanne, CH-1015 Lausanne, Switzerland
\item[\lyon] Institut de Physique Nucl\'eaire de Lyon, 
     IN2P3-CNRS,Universit\'e Claude Bernard, 
     F-69622 Villeurbanne, France
\item[\madrid] Centro de Investigaciones Energ{\'e}ticas, 
     Medioambientales y Tecnol\'ogicas, CIEMAT, E-28040 Madrid,
     Spain${\flat}$ 
\item[\florida] Florida Institute of Technology, Melbourne, FL 32901, USA
\item[\milan] INFN-Sezione di Milano, I-20133 Milan, Italy
\item[\moscow] Institute of Theoretical and Experimental Physics, ITEP, 
     Moscow, Russia
\item[\naples] INFN-Sezione di Napoli and University of Naples, 
     I-80125 Naples, Italy
\item[\cyprus] Department of Physics, University of Cyprus,
     Nicosia, Cyprus
\item[\nymegen] University of Nijmegen and NIKHEF, 
     NL-6525 ED Nijmegen, The Netherlands
\item[\caltech] California Institute of Technology, Pasadena, CA 91125, USA
\item[\perugia] INFN-Sezione di Perugia and Universit\`a Degli 
     Studi di Perugia, I-06100 Perugia, Italy   
\item[\peters] Nuclear Physics Institute, St. Petersburg, Russia
\item[\cmu] Carnegie Mellon University, Pittsburgh, PA 15213, USA
\item[\potenza] INFN-Sezione di Napoli and University of Potenza, 
     I-85100 Potenza, Italy
\item[\prince] Princeton University, Princeton, NJ 08544, USA
\item[\riverside] University of Californa, Riverside, CA 92521, USA
\item[\rome] INFN-Sezione di Roma and University of Rome, ``La Sapienza",
     I-00185 Rome, Italy
\item[\salerno] University and INFN, Salerno, I-84100 Salerno, Italy
\item[\ucsd] University of California, San Diego, CA 92093, USA
\item[\sofia] Bulgarian Academy of Sciences, Central Lab.~of 
     Mechatronics and Instrumentation, BU-1113 Sofia, Bulgaria
\item[\korea]  The Center for High Energy Physics, 
     Kyungpook National University, 702-701 Taegu, Republic of Korea
\item[\purdue] Purdue University, West Lafayette, IN 47907, USA
\item[\psinst] Paul Scherrer Institut, PSI, CH-5232 Villigen, Switzerland
\item[\zeuthen] DESY, D-15738 Zeuthen, Germany
\item[\eth] Eidgen\"ossische Technische Hochschule, ETH Z\"urich,
     CH-8093 Z\"urich, Switzerland
\item[\hamburg] University of Hamburg, D-22761 Hamburg, Germany
\item[\taiwan] National Central University, Chung-Li, Taiwan, China
\item[\tsinghua] Department of Physics, National Tsing Hua University,
      Taiwan, China
\item[\S]  Supported by the German Bundesministerium 
        f\"ur Bildung, Wissenschaft, Forschung und Technologie
\item[\ddag] Supported by the Hungarian OTKA fund under contract
numbers T019181, F023259 and T037350.
\item[\P] Also supported by the Hungarian OTKA fund under contract
  number T026178.
\item[$\flat$] Supported also by the Comisi\'on Interministerial de Ciencia y 
        Tecnolog{\'\i}a.
\item[$\sharp$] Also supported by CONICET and Universidad Nacional de La Plata,
        CC 67, 1900 La Plata, Argentina.
\item[$\triangle$] Supported by the National Natural Science
  Foundation of China.
\end{list}
}
\vfill


\clearpage

\end{document}